\documentclass[twocolumn, twocolappendix]{aastex631}

\usepackage{color}
\usepackage{url}
\usepackage{placeins}

\urldef{\urlA}\url{https://simple-bd-archive.org/load_solo/CWISE%20J210640.16+250729.0}

\begin{document}

\title{Sinking Silicates I: Characterizing the benchmark system containing the T0 brown dwarf CWISE
J210640.16+250729.0 using JWST}

\author[0000-0003-4083-9962]{Austin Rothermich}
\affiliation{Department of Astrophysics, American Museum of Natural History, New York, NY, USA}
\affiliation{Department of Physics, Graduate Center, City University of New York, 365 5th Ave., New York, NY 10016, USA}
\affiliation{Department of Physics and Astronomy, Hunter College, City University of New York, 695 Park Avenue, New York, NY, 10065, USA}
\affiliation{Backyard Worlds: Planet 9}

\author[0000-0001-6251-0573]{Jacqueline K. Faherty}
\affiliation{Department of Astrophysics, American Museum of Natural History, New York, NY, USA}

\author[0000-0003-4600-5627]{Ben Burningham}
\affiliation{Department of Physics, Astronomy and Mathematics, University of Hertfordshire, Hatfield, UK}

\author[0000-0002-0900-6076]{Catherine Manea}
\affiliation{Department of Physics \& Astronomy, University of Utah, Salt Lake City, UT 84112, USA}

\author[0000-0002-2682-0790]{Emily Calamari}
\affiliation{Department of Astrophysics, American Museum of Natural History, New York, NY, USA}

\author[0000-0002-2011-4924]{Genaro Su\'arez}
\affiliation{Department of Astrophysics, American Museum of Natural History, New York, NY, USA}

\author[0000-0001-8170-7072]{Daniella C. Bardalez Gagliuffi}
\affiliation{Department of Physics \& Astronomy, Amherst College, Amherst, MA, USA}

\author[0000-0002-9873-1471]{John M. Brewer}
\affiliation{Department of Physics \& Astronomy, San Francisco State University, 1600 Holloway Ave., San Francisco, CA 94132, USA}

\author[0000-0002-1821-0650]{Kelle L. Cruz}
\affiliation{Department of Physics and Astronomy, Hunter College, City University of New York, New York, NY, USA}
\affiliation{Department of Astrophysics, American Museum of Natural History, New York, NY, USA}
\affiliation{Department of Physics, Graduate Center, City University of New York, New York, NY, USA}

\author{Josefine Gaarn}
\affiliation{Department of Physics, Astronomy and Mathematics, University of Hertfordshire, Hatfield, UK}

\author[0000-0002-2592-9612]{Jonathan Gagn\'{e}}
\affiliation{Plan\'{e}tarium Rio Tinto Alcan, Montreal, Quebec, Canada}
\affiliation{D\'{e}partement de Physique, Universit\'{e} de Montr\'{e}al, Montreal, Quebec, Canada}

\author[0000-0003-4636-6676]{Eileen C. Gonzales}
\affiliation{Department of Physics \& Astronomy, San Francisco State University, 1600 Holloway Ave., San Francisco, CA 94132, USA}

\author[0000-0002-8871-773X]{Marina E. Gemma}
\affiliation{Department of Geosciences, Stony Brook University, Stony Brook, NY 11794, USA}
\affiliation{Department of Earth and Planetary Sciences, American Museum of Natural History, New York, NY 10024, USA}

\author[0000-0002-9511-0901]
{Vikt\'{o}ria Kecskem\'{e}thy}
\affiliation{Department of Physics, Astronomy and Mathematics, University of Hertfordshire, Hatfield, UK}

\author[0000-0002-5251-2943]{Mark S. Marley}
\affiliation{Lunar and Planetary Laboratory, University of Arizona, 1629 E. University Boulevard, Tucson, AZ 85721, USA}

\author[0000-0001-5610-5328]{Caprice L. Phillips}
\altaffiliation{NASA Sagan Fellow}
\affiliation{Department of Astronomy \& Astrophysics, University of California, Santa Cruz, CA 95064, USA}

\author[0000-0001-6627-6067]{Channon Visscher}
\affiliation{Department of Chemistry and Planetary Sciences, Dordt University, Sioux Center, IA, USA}
\affiliation{Center for Exoplanetary Systems, Space Science Institute, Boulder, CO, USA}

\author[0000-0003-0489-1528]{Johanna M. Vos}
\affiliation{School of Physics, Trinity College Dublin, The University of Dublin, Dublin, Ireland}

\author[0000-0001-8818-1544]{Niall Whiteford}
\affiliation{Department of Astrophysics, American Museum of Natural History, New York, NY, USA}



\begin{abstract}
In this study, we present the full (97.3\% complete) 0.8--12.5 $\mu$m spectral energy distribution (SED) of an L/T transition object, CWISE J210640.16+250729.0 (CW2106), using the James Webb Space Telescope (JWST). We provide a full characterization of the host star's elemental abundances and age. We empirically derive the bolometric luminosity ($L_{\rm bol}\approx-4.825$ $\textup{L}_\odot$) of CW2106, and obtain estimates of its mass (M$\approx50-62$ M$_{\rm Jup}$), radius (R$\approx0.83-0.87$ R$_{\rm Jup}$), effective temperature ($T_{\rm eff}$$\approx1213$ K), and surface gravity ($\log~g$$\approx5.28$ dex). We find the near-infrared (near-IR) spectrum ($0.8-2.5 ~\mu$m) is best reproduced with cloudy atmospheric models while the mid-infrared (mid-IR) spectrum ($5-12.5 ~\mu$m) is best reproduced with cloudless models. This suggests a cloud layer restricted to only the deepest observable parts of the atmosphere and is qualified by the lack of a 9 $\mu$m silicate feature. Making use of the Mg/Si ratio of the primary, alongside thermochemical models, we predict the clouds in CW2106 to be composed primarily of enstatite (MgSiO$_3$), removing $\sim23\%$ of the bulk oxygen out of the atmosphere. Future retrieval studies will be able to help investigate the existence and full impact of these cloud species.

\end{abstract}



\section{Introduction} \label{sec:intro}

Brown dwarfs, objects with masses below the hydrogen burning mass limit (HBML) ($\sim$78.5 M$_{Jup}$; \citealt{Chabrier_etal2023}), represent the lowest mass products of the stellar initial mass function (IMF) \citep{Kirkpatrick_etal2021, Kirkpatrick_etal2024, Best_etal2024}. While brown dwarfs may form in a similar process as stars \citep{Bate_etal2002}, the lack of hydrogen fusion in their cores results in a drastically different evolution than for objects whose masses are above the HBML. Unlike stars on the main sequence, which remain relatively stable over long periods of time, brown dwarfs slowly cool and contract as they age. As a consequence, the continually cooling atmospheres of brown dwarfs give rise to different molecular species and condensates as they age, morphing them from the warm ``L'' spectral types into the cooler T-types, and eventually to the cold Y spectral class \citep{Burrows_etal1997, Kirkpatrick_etal2005, Cushing2010}. 

A particularly interesting stage of brown dwarf evolution occurs during the phase in which they cool below L-dwarf  temperatures into that of the T-dwarfs, the aptly named L/T transition. This transition, which occurs over a narrow temperature range (T$_{eff} \approx1400 - 1100$ K; \citealt{Kirkpatrick_etal2000,Cushing_etal2008,Stephens_etal2009}), results in distinct observed differences in spectral morphology. Prominent methane (CH$_4$) bands appear in the J-, H-, and K-bands as CH$_4$ replaces carbon monoxide (CO) as the main carbon bearing molecule \citep{Burgasser_etal2002}. Silicate condensates (i.e. clouds), such as enstatite (MgSiO$_3$) and forsterite (Mg$_2$SiO$_4$), which form in the upper atmospheres of the warmer L-dwarfs \citep{2006asup.book....1L} sediment below the photosphere during this transition \citep{Suarez_etal2022}. Due in part to the difficulty in cloud modeling, no current self-consistent atmospheric or evolutionary models can fully and accurately reproduce the L/T transition (see \citealt{Morley_etal2024} for more discussion).

Obtaining precise fundamental parameters of L/T transition objects, such as mass (M), radius (R), effective temperature ($T_{\rm eff}$), and bolometric luminosity ($L_{\rm bol}$), is a crucial step in understanding this stage of brown dwarf evolution. Young (i.e. low surface gravity) brown dwarfs along the L/T transition, for example, appear to have cooler $T_{\rm eff}$'s and redder colors than their field-aged counterparts \citep[e.g.,][]{Metchev_2006, Luhman_2007}. Cloud sedimentation at the L/T boundary has been found to be less efficient in the younger objects as well \citep[e.g.,][]{Marley_2012, Suarez_2023}. The canonical approach to deriving these fundamental parameters is through the assemblage of broad wavelength coverage spectral energy distributions (SEDs) using flux calibrated optical to infrared spectrophotometry along with evolutionary models (c.f. \citealt{Stephens_etal2009,2015ApJ...810..158F, 2023ApJ...959...63S}). This method relies on having precise distances and ages, the main sources of error in the derivation of fundamental parameters.

While parallaxes for brown dwarfs are becoming more common, with $\sim$361 objects $\lesssim$25 pc having a trigonometric parallax \citep{Dupuy_etal2012,Faherty_etal2012,Bardalez_2019,Best_etal2020,Kirkpatrick_etal2021, 2021A&A...649A...6G}, ages for brown dwarfs are still difficult to obtain. The majority of brown dwarf ages come from objects belonging to young moving groups (e.g. \citealt{Liu_etal2013,Gagne_etal2014,Gagne_etal2015, Faherty_etal2016}), and are therefore limited to ages $\lesssim$1 Gyr. For isolated field brown dwarfs, which comprise the majority of the known brown dwarfs, there are no observational methods which can precisely constrain their ages. Instead, the field population requires the use of age estimates with extremely large error bars. This means for a typical brown dwarf, whose age is poorly constrained, estimates for mass, radius, $T_{\rm eff}$, and $\log~g$ obtained from evolutionary models are strongly impacted by the choice of age distribution. 

To illuminate this, we can take a look at the analyses performed by \cite{2015ApJ...810..158F} and \cite{2023ApJ...959...63S}, whose fundamental parameter samples contain 130 overlapping objects. The two studies made use of different age distributions for field objects: \cite{2015ApJ...810..158F} used a uniform distribution spanning 0.5--10 Gyrs, while \cite{2023ApJ...959...63S} implemented the age distribution from \cite{Dupuy_Liu2017}, a probability density function spanning 0.01--10 Gyr which prefers younger ages, leading to different fundamental parameters for $\sim$37 objects. 2MASS J03185403-3421292 (a field age L7; \citealt{Kirkpatrick_etal2008}) for example has differences in mass, radius, $T_{\rm eff}$, and $\log~g$ of $\sim$82\%, 36\%, 18\%, and 19\% respectively. Many of the 37 objects with differences between the two studies are likewise field objects, whose parameters are similarly impacted by the change in age distribution choice. 

Benchmark brown dwarfs-- defined here-in as objects that are co-moving with a higher mass star-- are an ideal sample for deriving fundamental parameters (e.g \citealt{Rothermich24}). The coevality of the pair means that many of the parameters derived for the host star, such as distance, age, and metallicity, are also applicable for its brown dwarf companion. Host star parallaxes are widely available thanks to all sky astrometry programs such as \textit{Gaia} \citep{Gaia}, providing distances for their brown dwarf companions without the need for observationally expensive parallax programs. Many host stars are also ideal for age dating, either through comparisons with stellar isochrone models (e.g. \citealt{Choi_etal2016}), white dwarf cooling models \citep[e.g.][]{Wood1995,Fontaine_2001,Renaldo2010, Kiman_etal2022}, or gyrochronology \citep[e.g.][]{Skumanich1972,Mamajek2008,Popinchalk_etal2021}, all of which make it possible to derive well constrained ages $>$1 Gyr, something not currently possible for isolated brown dwarfs. The fundamental parameters derived for these benchmarks are therefore free from the errors associated with probabilistic age distribution estimates, resulting in more precise estimates of mass, radius, $T_{\rm eff}$, and $\log~g$ than otherwise possible for typical field objects. 

Studying benchmark systems provides another advantage over isolated field objects in that the elemental abundances of the system can be constrained through the host star, particularly if it is a solar-like F-, G-, or K-type star for which these analyses are best suited. Assuming the two objects formed out of the same natal environment, presumably the pair will share the same primordial makeup. Indeed, studies comparing the abundances between stellar binary twins, pairs in which $\Delta$T$_{eff}<$200K, have found that $\sim$80\% of wide systems (300-50,000 AU) share a similar chemical makeup within $\sim$0.02 dex \citep{Hawkins_2020,Teske_2024}. This trend appears to hold even in the widest pairs, with $\sim$70\% of wide co-moving stellar twins ($\lesssim100,000$ AU) being chemically homogeneous \citep[e.g.,][]{Nelson_etal2021, Yong_etal2023}. Brown dwarf benchmark systems share a common formation pathway with typical stellar binaries \citep[e.g.,][]{Bate_2009a, Bate_2012}, and as such are also expected to be chemically similar to their stellar relative. 

Newer approaches such as spectral inversion techniques, otherwise known as retrievals, are now able to model a brown dwarf's atmospheric gas abundances, $T_{\rm eff}$, cloud properties, and other fundamental parameters \citep{Line_etal2015,Gonzales_etal2020,Burningham_etal2021, Zalesky_etal2022, Vos_etal2023,Lew_etal2024,Kothari_etal2024}. However, properties such as atmospheric gas abundances are still difficult to model, relying on sometimes incomplete or incorrect absorption cross sections \citep{Burningham_etal2017}. Benchmark brown dwarfs, with elemental abundances known \textit{a priori} which can be used to ground the analysis, consequently represent ideal targets for retrieval studies. While retrievals are able to estimate the atmospheric abundances of a brown dwarf, the abundances provided for benchmark systems via their host stars represent the bulk composition. These bulk abundances can be used, for example, along with retrieved abundances to investigate oxygen sequestration due to cloud condensates (c.f. \citealt{Calamari_etal2022, Wang_etal2022, Gaarn_etal2023}), or to predict the cloud species expected to form \citep{Calamari_etal2024, 2024ApJ...972..172P}.

In this paper, we present the first SED and fundamental parameters of CWISE J210640.16+250729.0, an L-T transition companion to a K-dwarf. In Section \ref{sec:Overview} we present an overview of the discovery and properties of the system. Section \ref{sec:obs} describes the observations and data reduction carried out for both the brown dwarf and its host star. We characterize the host star in Section \ref{sec:host_characterization}, including an analysis of its elemental abundances and age determination. In Section \ref{comp_characterization} we assign a spectral type to CWISE J210640.16+250729.0 and derive its fundamental parameters using its full SED. Section \ref{sec:spec_analysis} presents a deeper look into the spectral features observed in the SED of CWISE J210640.16+250729.0, while Section \ref{sec:modeling} compares its SED with forward model grids. In Section \ref{sec:clouds} we discuss the expected species of silicate clouds using recent thermochemical models. The results of the forward model comparisons are placed into context with the system's known properties in Section \ref{sec:benchmarkTime!}. Our conclusions are presented in Section \ref{sec:conclusions}.

\section{System Overview} \label{sec:Overview}

CWISE J210640.16+250729.0 (hereafter referred to as CW2106) was found to be a co-moving companion to the K6V dwarf BD+24 4329 (BD+24 from this point on) in \cite{Rothermich24}. After initially identifying the pair visually through the citizen science project Backyard Worlds: Planet 9 (\citealt{2017ApJ...841L..19K}),  \cite{Rothermich24} used the Bayesian algorithm \texttt{CoMover} \citep{2021ascl.soft06007G} to assess the likelihood of the pair being physically associated using each object's coordinates, proper motions, and parallax (for CW2106, its photometric distance estimate), and their associated errors. They found the system had a 99.4\% probability of being associated, above the 90\% cutoff used to determine ``high likelihood'' systems. The pair have an angular separation of 1124 arcsec ($\sim$18.7 arcmin), translating to a projected physical separation of $\sim$38,000 AU. While no spectrum of CW2106 was presented, CW2106 was assigned an estimated phototype of $\sim$T2. 

\begin{figure}
    \centering
    \includegraphics[width=1.0\linewidth]{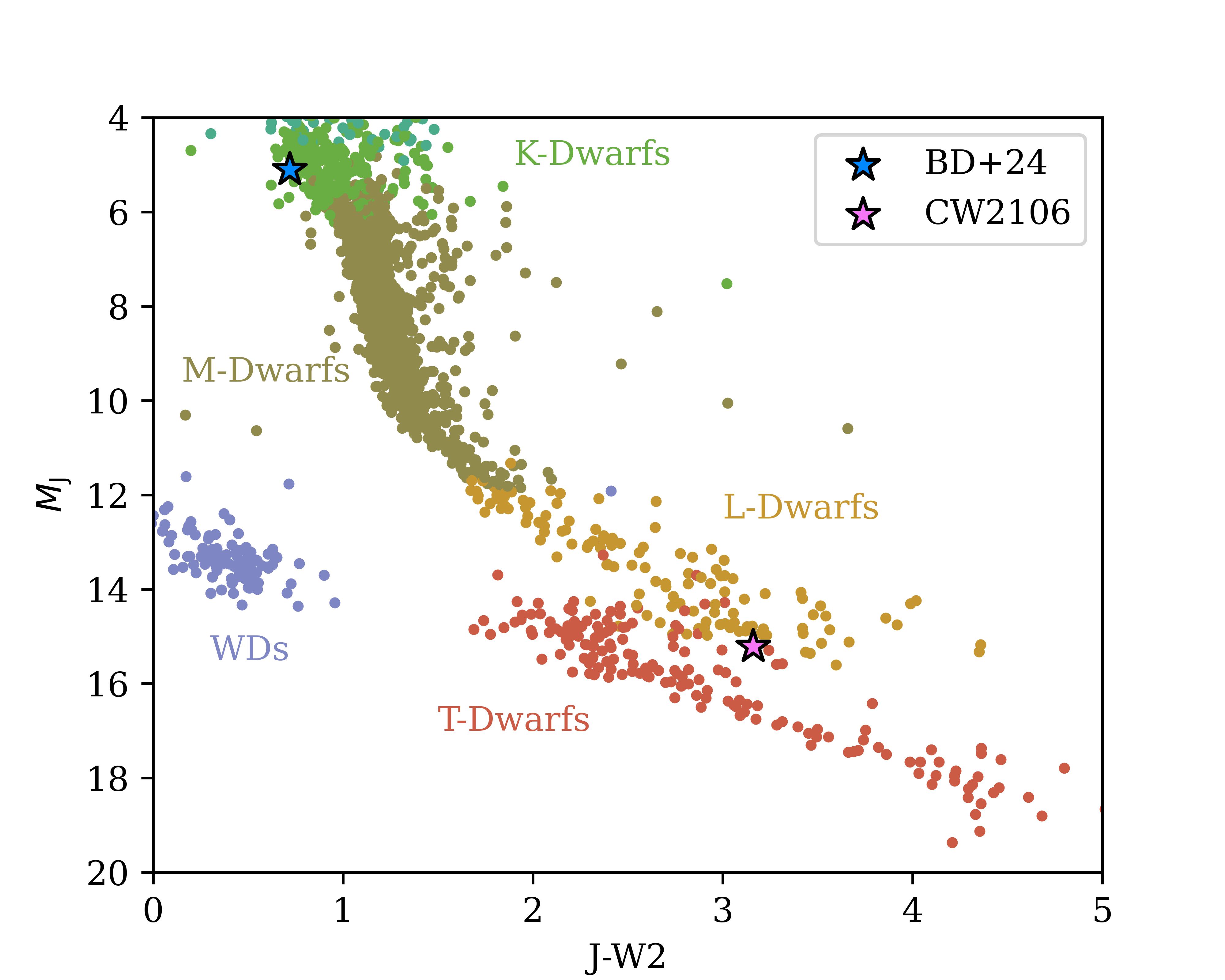}
    \caption{Color-magnitude diagram for the 20 pc parallax sample of \cite{Kirkpatrick_etal2024}, with the different spectral types colored and labeled. BD+24 is shown as the blue pointed star, with CW2106 shown as a pink pointed star.} \label{fig:cmd}
\end{figure}

On a Gaia color-magnitude diagram (CMD), BD+24 has a normal brightness compared to objects of a similar color. However, it stands out a bit more when using infrared colors as shown in the CMD in Figure \ref{fig:cmd}, with BD+24 being slightly under-luminous in J for its J-W2 color. In contrast, CW2106 appears to be under-luminous for its J-W2 color in Figure \ref{fig:cmd}. CW2106 was estimated to have a mass of $\sim36$  M$_{Jup}$ \citep{Rothermich24}, giving the system a mass ratio of 0.05, a binding energy of $E_B=0.1\times10^{41}$ ergs, and a dynamical dissipation lifetime of 4 Gyrs. A summary of the BD+24 \& CW2106 system properties are summarized in Table \ref{tab:sys_prop}.

\begin{deluxetable*}{lccccc}
    \tabletypesize{\scriptsize}
    \tablewidth{0pt}
    \tablenum{1}
    \tablecolumns{3}
    \tablecaption{BD+24 4329 \& CW2106 System Properties.\label{tab:sys_prop}}
    \tablehead{ 
    \colhead{Property} & 
    \colhead{BD+24 4329} &
    \colhead{Ref} &
    \colhead{CW2106} &
    \colhead{Ref}}
    \startdata
        SpType & K6V & Steph1986 & T0$\pm$1 & This Work \\
        RA (deg) & 317.00781 & DR3 & 316.66735 & CW2020 \\
        Dec (deg) & 25.17551 & DR3 & 25.12471 & CW2020 \\
        PMRA (mas yr$^{-1}$) & -21.89$\pm$0.01 & DR3  & -39.5$\pm$14.7 & CW2020 \\
        PMDec (mas yr$^{-1}$)& -162.53$\pm$0.01 & DR3 & -132.6$\pm$13.3 & CW2020  \\
        RV (km s$^{-1}$) & -5.31$\pm$0.15 & DR3 & --- & ---  \\
        Vsin(i) (km s$^{-1}$) & 6.02$\pm$2.19 & DR3  & --- & --- \\
        Plx (mas) & 29.31$\pm$0.01 & DR3  & --- & ---  \\
        \cutinhead{Photometry}
        g$_{PS1}$ (mag) & 10.21$\pm$0.04 & PS1 & --- & --- \\
        r$_{PS1}$ (mag) & 10.11$\pm$0.01 & PS1 & --- & --- \\
        i$_{PS1}$ (mag) & 9.53$\pm$0.15 & PS1 & --- & --- \\
        z$_{PS1}$ (mag) & 10.08$\pm$0.01 & PS1 & 21.19$\pm$0.08 & PS1 \\
        y$_{PS1}$ (mag) & 9.97$\pm$0.09 & PS1 & 19.85$\pm0.16$ & PS1 \\
        J (mag) & 7.77$\pm$0.02 & 2MASS & 17.88$\pm$0.05 & UHS \\
        H (mag) & 7.19$\pm$0.03 & 2MASS & --- & --- \\
        K (mag) & 7.06$\pm$0.03 & 2MASS & --- & --- \\
        W1 (mag) & 6.92$\pm$0.01 & CW2020 & 15.33$\pm$0.02 & CW2020 \\
        W2 (mag) & 7.05$\pm$0.01 & CW2020 & 14.72$\pm$0.02 & CW2020 \\
        W3 (mag) & 7.04$\pm$0.02 & AllWISE & 12.56$^*$ & CW2020 \\
        W4 (mag) & 7.07$\pm$0.10 & AllWISE & 8.83$^*$ & CW2020 \\
        \cutinhead{Fundamental Parameters}
        $L_{\rm bol}$ (\(\textup{L}_\odot\)) & -0.729$\pm$0.039 & Tuch2024 & -4.825$\pm$0.005 & This Work \\
        $T_{\rm eff}$ (K) & 4511$\pm$21 & This Work & 1213$\pm21$ & This Work \\
        Mass (\(\textup{M}_\odot\), M$_{Jup}$) & 0.71$\pm$0.04  & Tuch2024  & 56$\pm6$ & This Work\\
        Radius (\(\textup{R}_\odot\), R$_{Jup}$) & 0.71$\pm$0.04 & Tuch2024 & 0.85$\pm0.02$ & This Work \\
        $\log~g$ (dex) & 4.42$\pm$0.05 & This Work & 5.28$\pm0.07$ & This Work \\
        Age (Gyrs) & 3.3$^{+0.7}_{-0.5}$ & This Work & --- & --- \\
    \enddata
    \tablecomments{\textbf{References} - DR3, \cite{GaiaDR3}; CW2020, \cite{CatWISE2020}; 2MASS, \cite{2MASS}; PS1, \cite{Panstarrs}; AllWISE, \cite{AllWISE2014} Tuch2024, \cite{Tuchow2024}; Steph1986, \cite{Stephanson1986}; UHS, \cite{UHSsurvey}.\newline
    a - The probability of the pair being physically associated. \newline
    * - Denotes magnitude limit.}
\end{deluxetable*}

\section{Observations} \label{sec:obs}
In this section we present new data for the CW2106 system. All data is available from the SIMPLE Archive\footnote{\urlA\ }. 
\subsection{Magellan/FIRE} \label{sec:fire_spec}
\begin{figure*}
    \centering
    \includegraphics[width=1.0\textwidth]{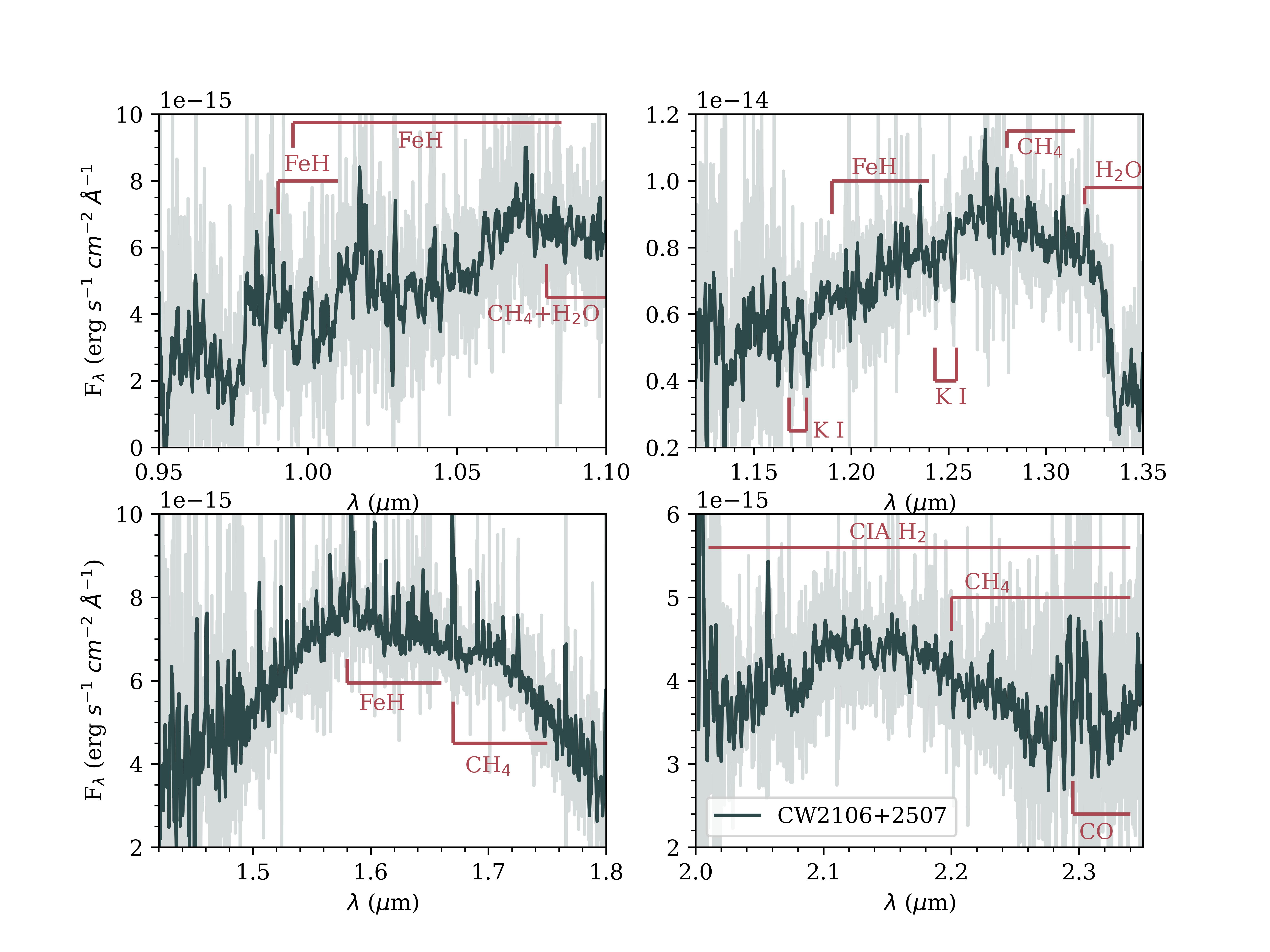}
    \caption{Magellan FIRE/Echelle spectrum of CW2106 shown in grey, with the smoothed spectrum over-plotted in dark green. Major absorption features have been labeled in each panel.} \label{fig:fire_spectrum}
\end{figure*}

CW2106 was observed with the Folded-port Infrared Echellette (FIRE, \citealt{FIRE}) on the Magellan Baade telescope, located at the Las Campanas observatory. Near infrared data (0.82 -- 2.51 $\mu$m) was taken on 28 May 2024 with the instrument in echelle mode, providing a moderate resolving power of $\lambda/\Delta\lambda$ $\approx$ 6,000. Light cirrus was present throughout the observations, with seeing between $\sim$0.7 - 1.0 arcsec and at an airmass of 1.79. The telluric star HIP 92177 was observed at an airmass of 2.08 after the science observation. Data were reduced using the FIREHOSE package, which is based on the MASE and SpeX reduction tools \citep{Vacca_etal2003, Cushing_etal2004, Bochanski_etal2009}. The final FIRE spectrum is shown in Figure \ref{fig:fire_spectrum}

\subsection{JWST}\label{sec:obs-jwst}

\subsubsection{Data Acquisition}

CW2106 was observed with The James Webb Space Telescope (\textit{JWST}, \citealt{2023PASP..135d8001R}) on 24 June 2024 as part of the cycle 2 GO program 3670 (PI: Burningham). Observations were performed with the Near-Infrared Spectrograph (NIRSpec, \citealt{2022A&A...661A..80J}) and Mid-Infrared Instrument (MIRI, \citealt{2015PASP..127..584R}). 

NIRSpec was used in low resolution prism mode with the S200A1 fixed slit to obtain near infrared data (0.6-5.3 $\mu$m) with a resolving power of $\lambda/\Delta\lambda$ $\approx$ 100. Target acquisition was carried out using WATA method, clear filter, subarray 32 and NRSRAPID readout pattern. The science observation was performed with 10 groups per integration, 1 integration per exposure, and 3 dithers for a total integration time of 51.475 s. The resulting spectrum has an average signal-to-noise (SNR) of $\sim52$.

MIRI was used to obtain mid infrared data (5-14 microns) low-resolution spectrometer (LRS) in fixed-slit mode providing a resolving power of $\lambda/\Delta\lambda$ $\approx$ 100 at 7.5 $\mu$m. Target acquisition was completed using the F1000W filter and FAST readout pattern. Science observations were taken with the FASTR1 readout pattern, 100 groups/integration, 1 integration/exposure and  2 dithers resulting in a total exposure time of 555.008 s, resulting in an average SNR of $\sim72$ across the full wavelength coverage.  

\subsubsection{Data Reduction}
Data from both NIRSpec and MIRI were reprocessed using version 1.15.1 of the official JWST pipeline \citep{bushouse_2022_7229890} using Calibration Reference Data System context file jwst\_1322.pmap, with default parameters. We started from the unprocessed raw data downloaded from the Mikulski Archive for Space Telescopes (MAST). Corrections for cosmic rays, dark currents, count rate non-linearity, bad pixel flagging, and other detector level corrections are applied to the raw data during stage 1 using the ``calwebb\_detector1'' pipeline. The ``calwebb\_spec2'' pipeline in stage 2 is then used with the data output from stage 1 to produce calibrated individual exposures. The individual exposures are combined in stage 3, where the final spectrum is extracted from the combined image. The final reduced data is presented in Section \ref{comp_characterization}.  All JWST data used in this paper can be found in MAST: \dataset[10.17909/rgey-0312]{http://dx.doi.org/10.17909/rgey-0312}.

%\subsection{LBT/PEPSI} \label{sec:pepsi_spec}
%BD+24 4329 was observed on \textcolor{red}{Observation Date} with the Potsdam Echelle Polarimetric and Spectroscopic Instrument (PEPSI, \textcolor{red}{REFERENCE}), located on the Large Binocular Telescope (LBT, \textcolor{red}{REFERENCE}) in Safford, AZ. For the blue arm observations, cross-disperser 3 was used to obtain optical data from 4800 -- 5441 \AA, while red arm observations were performed with cross-disperser 6 to obtain data from 7419 -- 9140 \AA. Observations were carried out in binocular mode with fiber 200, with exposure times for cross-dispersers 3 and 6 of 2800s and 250s respectively, providing an average resolution of 130,000 and SNR of $>$200 across both cross-dispersers. \textcolor{red}{Add reduction info here.}

\subsection{Tull Coude Spectrograph}\label{sec:tull_spec}

\begin{figure*}
    \centering
    \includegraphics[width=1.0\textwidth]{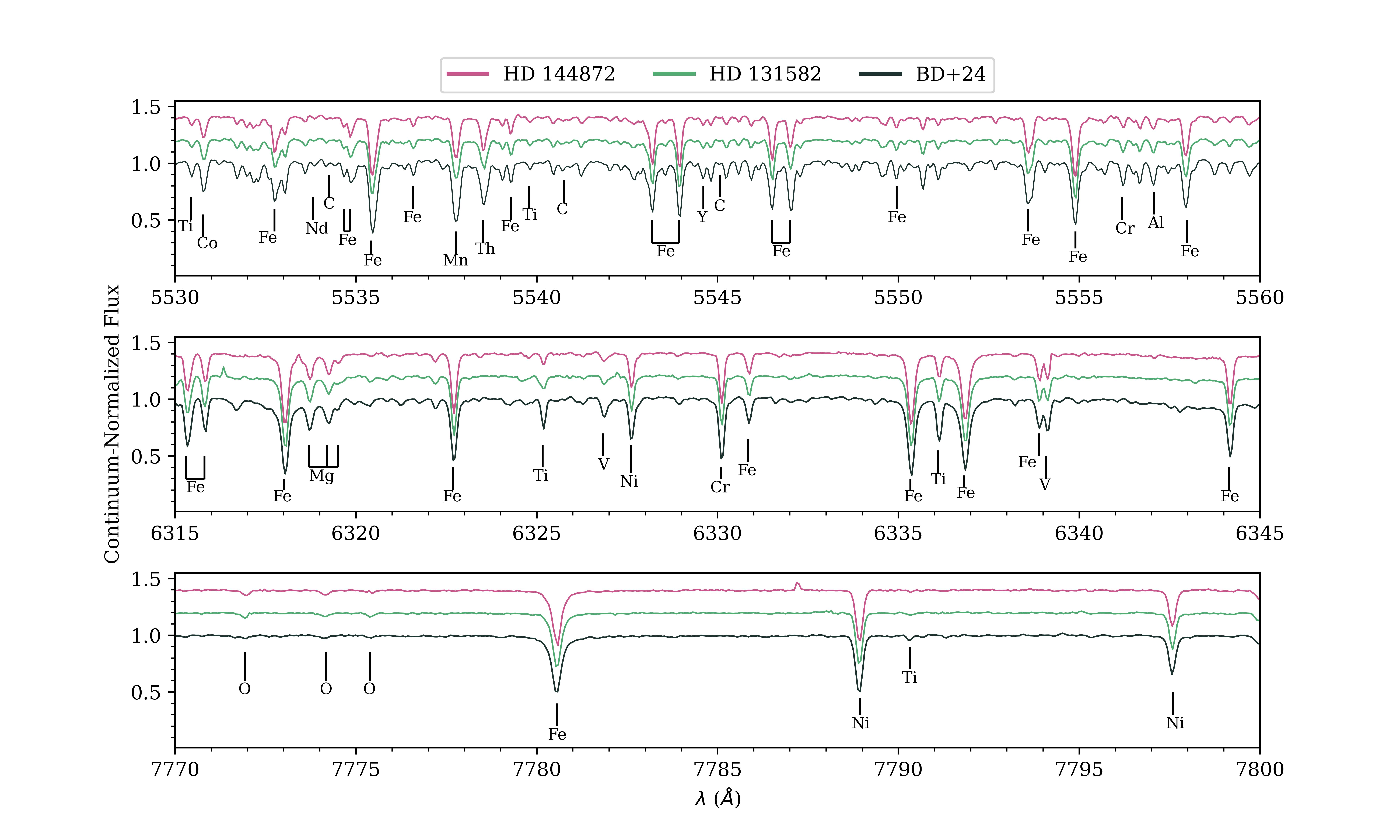}
    \caption{Subsets of BD+24's spectrum, as described in Section \ref{sec:tull_spec}, in black along with the spectra of HD 131582 and HD 144872 as pink and green lines respectively, all separated by a constant. Prominent absorption lines are labeled.}
    \label{fig:BD24spectrum}
\end{figure*}

We observed BD+24 twice, first on July 13, 2024 and again on July 24, 2025 with the 2.7 m  Harlan J. Smith Telescope, located at McDonald Observatory in Fort Davis, TX. Two stars from the stellar abundance catalog presented in \cite{Brewer_cat}, HD 131582 (spectral type K3) and HD 144872 (spectral type K3), were observed as abundance standards on April 12, 2025. Both abundance standards are similar brightnesses and colors as one another (V=8.608 mag, B-V=0.993 mag and V=8.61 mag, B-V=0.96 mag for HD 131582 and HD 144872 respectively), but slightly brighter and bluer than BD+24 (V=9.88 mag, B-V=1.16 mag). We used the Tull Coude Spectrograph in TS23 mode, with slit \#4, grating E2, and the TK3 detector, resulting in continuous spectral coverage from 3500--10500 \AA\ at a resolution of R$\sim$60,000, achieving an average SNR of $\sim200$. The raw Tull Coudé Spectrograph data are initially processed using the Tull Coudé Spectrograph Data Reduction Pipeline (\texttt{TSDRP}\footnote{\url{https://github.com/grzeimann/TSDRP}}). This pipeline performs essential calibration and extraction steps, including bias subtraction, trace identification, scattered light subtraction, wavelength calibration, flat-field correction, cosmic ray rejection, and spectral extraction for each spectral order. Additionally, TSDRP provides deblazing, continuum normalization, and order combination to produce a single, fully processed spectrum for each exposure.  All spectra were then corrected for radial velocity shifts using \texttt{iSpec}'s cross-correlation function with a NARVAL Solar spectrum provided with the installation \citep{2014A&A...569A.111B}. Finally, all spectra of the same object were coadded, weighting by the signal-to-noise of each exposure. A portion of the spectrum of BD+24 is shown in Figure \ref{fig:BD24spectrum}.

\section{BD+24 4329 Characterization} \label{sec:host_characterization}
The power of a benchmark system lies in utilizing well determined information on its host star. To properly utilize CW2106 as a benchmark brown dwarf, it is crucial to fully understand its host, BD+24 4329. In this section we provide a full characterization of BD+24 4329, estimating its age and deriving its elemental abundances. 

\subsection{Fundamental Parameters and Elemental Abundances} \label{ses:abundances}

\subsubsection{\texttt{BACCHUS}} \label{sec:bacchus}

We used version 72 of the Brussels Automatic Code for Characterizing High accUracy Spectra (\texttt{BACCHUS}, \citealt{2016ascl.soft05004M}) to derive the fundamental parameters and elemental abundances for BD+24 and the two abundance standards from \citet{Brewer_cat}. BACCHUS is a line-by-line spectral fitting code that synthesizes spectra using the radiative transfer code TURBOSPECTRUM \citep{Plez2012} adopting the MARCS model atmosphere grid \citep{Gustafsson2008} and assuming one-dimensional local thermodynamic equilibrium (1D LTE). Stellar parameters ($T_{\rm eff}$, $\log~g$, [M/H], microturbulent velocity,$v_{mic}$) are determined using Fe ionization-excitation balance, where \texttt{BACCHUS} iterates until it finds the combination of parameter values which results in an agreement between the abundances measured from Fe I and Fe II lines and null trends between each Fe I line's measured abundance both the excitation potential and reduced equivalent width of that transition.  Abundances are determined in a line-by-line way.  For each line of each element, BACCHUS synthesizes five absorption lines with abundances that range from -0.6 dex to +0.6 dex centered on the expected abundance of the line assuming a Solar abundance pattern scaled to the metallicity of the star.  The abundance of each line is then determined using four methods (the $\chi^2$, equivalent width, synthetic spectrum, and line intensity methods, see detailed description in \citealt{BAWLAS}), and each method returns an associated flag reporting the quality of the measurement.  Users can then include or exclude specific lines from their analysis by applying those flags. We adopt version 5 of the \textit{Gaia}-ESO linelist for our atomic transition data \citep{Heiter2021} and combine molecular transition data from numerous sources: CH from \citealt{Masseron2014}, C2, CN, OH, and MgH from T. Masseron, private communication, SiH from \citealt{Kurucz1992}, and TiO, FeH, and ZrO from B. Pelz, private communication.

The choice of instrument, spectral fitting technique, model atmospheres, atomic data, and lines being measured all affect the parameters and abundances determined for a star \citep[e.g.,][]{Jofre2019}.  A major goal of this work is to compare BD+24's parameters and abundances to those of typical disk stars in the \citet{Brewer_cat} catalog.  As such, we must adopt the same frame of reference as \citet{Brewer_cat}.  To achieve this, rather than performing standard abundance analysis, which can, again, lead to global offsets between our results and those of \citet{Brewer_cat} due to differences in methodology and data, we leverage the strictly line-by-line \textit{differential} technique, anchoring ourselves to the two abundance standards also observed by \citet{Brewer_cat} and thus the same parameter and abundance reference frame.  This method has been used extensively in the literature to determine precise parameters and abundances for stars \citep[e.g.,][and references therein]{Melendez2012, Liu2014, Galarza2016, Reggiani2017, Bedell2018, Spina2018, McKenzie2022, Yong_etal2023}.  Strictly line-by-line differential analysis essentially determines how \textit{different} a star is in e.g., $T_{\rm eff}$, $\log~g$, [M/H],$v_{mic}$, and detailed elemental abundance ([X/H]) with respect to a chosen reference star.  If we adopt the parameters and abundances reported by \citet{Brewer_cat} for the two abundance standards, then we can add to them our differential results for BD+24 to get parameters and abundances in the same reference frame as \citet{Brewer_cat}, enabling direct comparison to their catalog while minimizing systematic offsets.

To determine line-by-line differential parameters and abundances for BD+24 with respect to our two abundance standards HD 131582 and HD 144872, we must first measure their line-by-line abundances. We adopt the stellar parameters reported by \citet{Brewer_cat} (see ``Lit." column of Appendix Table \ref{tab:Brewer_abunds}) and use BACCHUS to measure Fe I, Fe II, and 24 other elemental abundances, only considering lines that the Gaia-ESO team report to be unblended in the Sun and Arcturus and have reliable atomic data ( \texttt{synflag} = Y and \texttt{gfflag} = Y) and also that BACCHUS returns as unflagged in all four abundance measurement methods.  Next, we solve for BD+24's parameters, once with respect to HD 131582 and again with respect to HD 144872, by performing differential Fe ionization-excitation with BACCHUS.  Differential Fe ionization-excitation is similar to standard Fe ionization-excitation except that it requires a null trend between lines' excitation potentials and the line-by-line Fe I abundance \textit{differences} between the star and its reference star. Similarly, it requires an agreement between the line-by-line Fe I and Fe II abundance differences between the star and its reference.  Finally, we determine line-by-line abundance differences between BD+24 and the two abundance standards among all shared lines between the spectra. 

Our approach returns two sets of differential abundances for BD+24, each of which is determined by taking the average of the line-by-line abundance differences between BD+24 and the reference star in each element and adding to them the abundances of the reference star reported by \citet{Brewer_cat}.  Associated differential abundance uncertainties are determined by taking the quadratic sum of the standard deviation in the line-to-line abundance differences and the impact of each parameter's uncertainty on the resulting abundance.\footnote{We determine the impact of each stellar parameter uncertainty on the resulting abundance by perturbing the model atmosphere by $\pm$ the parameter uncertainty and re-measuring abundances.} Final elemental abundances for BD+24 are taken as the average of the abundances determined with respect to each reference star, weighted by the associated uncertainties.  To validate our differential method, we also determine differential parameters and abundances for each reference star with respect to the other reference star.  The results of this validation are presented in Appendix \ref{sec:brewer_results}.
%Fundamental parameters are first derived by \texttt{BACCHUS} using the excitation-ionization balance technique. Starting from an initial guess, \texttt{BACCHUS} iterates until it finds the combination of $T_{\rm eff}$, $\log~g$, [M/H], and microturbulent velocity ($v_{mic}$) which results in an agreement between the abundances measured from Fe I and Fe II lines while avoiding correlations between each Fe I line's measured abundance and the excitation potential of that transition. Upon constraining stellar parameters, we use \texttt{BACCHUS} to determine abundances in 26 elements. For each line of each element, BACCHUS synthesizes five absorption lines with abundances that range from -0.6 dex to +0.6 dex centered on the expected abundance of the line assuming a Solar abundance pattern scaled to the metallicity of the star.  The abundance of each line is determined using four methods (the $\chi^2$, equivalent width, synthetic spectrum, and line intensity methods, see detailed description in \citealt{BAWLAS}).  Final elemental abundances are taken as the mean abundance determined from all lines where the four methods returned reliable (unflagged) abundance determinations.  Abundance uncertainties are reported as the standard error on the mean line-by-line abundance. Our final elemental abundances and stellar parameters derived with \texttt{BACCHUS} are presented in Table \ref{tab:abund_results}.

\begin{deluxetable}{lccc}[ht!]
    \tabletypesize{\scriptsize}
    \tablewidth{0pt}
    \tablenum{2}
    \tablecolumns{4}
    \tablecaption{BD+24 Derived Abundances and Parameters.}\label{tab:abund_results}
    \tablehead{ 
    \colhead{Property} & 
    \colhead{$\Delta_{131}^a$}&
    \colhead{$\Delta_{144}^b$}&
    \colhead{Final}} 
    \startdata
        $T_{\rm eff}$ (K) & 4500$\pm$79 & 4512$\pm$21 & 4511$\pm$21   \\\relax
        $\log~g$ (dex)  & 4.39$\pm$0.05 & 4.50$\pm$0.10 & 4.42$\pm$0.05   \\\relax
        [M/H] (dex)  & 0.00$\pm$0.26 & 0.00$\pm$0.06 & 0.00$\pm$0.06  \\\relax
        [Fe/H] (dex)  & -0.05$\pm$0.01 & -0.01$\pm$0.01 & -0.03$\pm$0.01  \\\relax
        [C/H] (dex)  & 0.17$\pm$0.02 & 0.16$\pm$0.02 & 0.16$\pm$0.02  \\\relax
        [N/H] (dex)  & -0.25$\pm$0.05 & -0.20$\pm$0.05 & -0.23$\pm$0.04  \\\relax
        [O/H] (dex)  & 0.11$\pm$0.08 & 0.09$\pm$0.09 & 0.11$\pm$0.06  \\\relax
        [Mg/H] (dex)  & 0.11$\pm$0.03 & 0.08$\pm$0.01 & 0.09$\pm$0.01  \\\relax
        [Si/H] (dex)  & 0.12$\pm$0.07 & 0.19$\pm$0.06 & 0.16$\pm$0.04 \\\relax
        [Ca/H] (dex)  & 0.03$\pm$0.03 & 0.00$\pm$0.03 & 0.02$\pm$0.02  \\\relax
        [Na/H] (dex)  & -0.14$\pm$0.10 & --- & -0.14$\pm$0.10  \\\relax
        [Al/H] (dex)  & -0.03$\pm$0.01 & -0.09$\pm$0.01 & -0.06$\pm$0.01  \\\relax
        [Ti/H] (dex)  & -0.08$\pm$0.04 & -0.05$\pm$0.02 & -0.06$\pm$0.02  \\\relax
        [V/H] (dex)  & -0.02$\pm$0.10 & -0.07$\pm$0.01 & -0.07$\pm$0.01  \\\relax
        [Cr/H] (dex)  & -0.08$\pm$0.01 & -0.09$\pm$0.01 & -0.09$\pm$0.01 \\\relax
        [Mn/H] (dex)  & -0.09$\pm$0.10 & -0.12$\pm$0.10 & -0.11$\pm$0.07  \\\relax
        [Ni/H] (dex)  & -0.02$\pm$0.02 & -0.02$\pm$0.02 & -0.02$\pm$0.01  \\\relax
        [Y/H] (dex)  & --- & -0.03$\pm$0.10 & -0.03$\pm$0.10  \\\relax
        C/O  & 0.63$\pm$0.08 & 0.64$\pm$0.09 & 0.63$\pm$0.06  \\\relax
        Mg/Si  & 1.00$\pm$0.08 & 0.80$\pm$0.06 & 0.87$\pm$0.04  \\
    \enddata
    \tablecomments{a - Differential values with respect to HD 131582. \newline
    b - Differential values with respect to HD 144872.}
\end{deluxetable}

\subsubsection{BD+24 Results} \label{sec:CW2106_abunds}

\begin{figure*}
    \centering
    \includegraphics[width=1.0\textwidth]{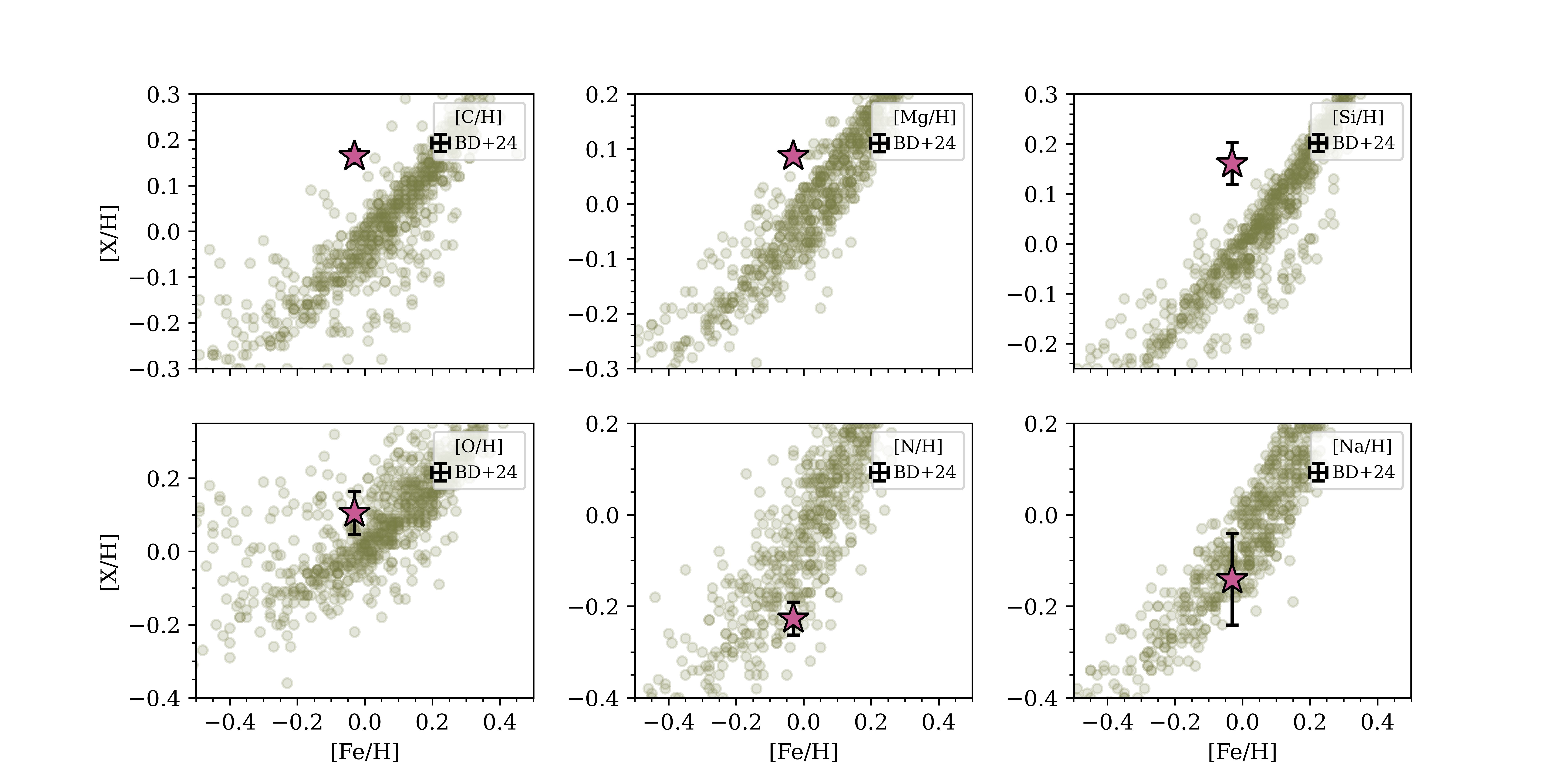}
    \caption{Abundances of select elements vs [Fe/H] for BD+24 (red star and error bars) along with the abundances of stars from \cite{Brewer_cat} (green points).}
    \label{fig:Abund_Comp_CW2106}
\end{figure*}

Figure \ref{fig:Abund_Comp_CW2106} shows the abundances of BD+24 for several elements compared with stars from the Brewer catalog. Overall, the abundances of BD+24 are in decent agreement with the distribution seen in the Brewer Catalog. BD+24 does show evidence of slight enhancement in several elements compared to other solar metallicity objects, most notably [C/H], [Mg/H], and [Si/H] which lie above the bulk of the population. The [O/H] of BD+24 also appears to be slightly elevated, though the Brewer catalog shows a larger degree of scatter in [O/H] for a given [Fe/H], with BD+24 fitting well with the observed scatter. In contrast, the observed [N/H] of BD+24 falls on the lower end of the distribution, suggesting a possible minor depletion in N.

The C/O ratio of BD+24 is found to be more carbon rich than the bulk of the Brewer Catalog with C/O=$0.63\pm0.06$, where the median C/O of the Brewer Catalog is $\sim0.47$ with a maximum of $0.66\pm0.07$ \citep{Brewer_etal2016}. The Mg/Si ratios of BD+24, in contrast, is slightly lower than the average Mg/Si of the Brewer Catalog, with Mg/Si=$0.87\pm0.04$ (compared with the median Mg/Si of 1.02 for the Brewer Catalog). However, \citep{Brewer_etal2016} note that the distribution of Mg/Si is quite large, with $\sim40\%$ of the sample having an Mg/Si$<1$.

\subsection{System Age} \label{sec:age}
In this section, we investigate multiple age indicators for BD+24, including kinematics, placement on CMDs, gyrochronology, and isochrones. 
  
Using the full kinematics from Gaia DR3, we investigated any possibility of BD+24 belonging to a young moving group (YMG) using Banyan $\Sigma$ \citep{2018ApJ...856...23G}. Banyan $\Sigma$ is a Bayesian algorithm which accepts as inputs the sky position, proper motions, radial velocity, and parallax of a source, and compares its position and space velocities with those of 27 known YMGs. The output is a list containing the probability of belonging to each of the YMGs tested, as well as belonging to the field population (defined here as not belonging to any of the 27 YMGs). Using the values listed in Table \ref{tab:sys_prop}, we get a probability of 99.9\% of BD+24 4329 belonging to the field population, with all 27 YMGs having a probaility of 0\%. BD+24's placement on the Gaia CMD is in agreement with these results, as it appears to have an average luminosity for typical stars of its color, rather than over-luminous as is typically seen with younger stars. 

The kinematics of BD+24 can also be used to calculate its projected orbit within the Milky Way. To do this we used the open-source code \texttt{galpy} \citep{2015ApJS..216...29B}, assuming a solar radius of 8 kpc and disk rotation speed of 220 km/s, and using the \texttt{MWPotential2014} potential as described in \cite{2015ApJS..216...29B}. This results in an orbit for BD+24 4329 with a periapsis of 7.16 kpc, an apoasis of 8.72 kpc, a maximum height above the plane of 0.07 kpc, and an eccentricity of 0.1 - all of which are consistent with an object belonging to the thin disk population, which has a scale height of z=300$\pm$50 pc \citep{Bland2016} and an eccentricity distribution which peaks at e$\sim$0.12 \citep{Yan2019}. 

With no strong age constraints provided by the previously discussed age diagnostics, we used a hybrid dating technique which incorporates both isochrone fitting and gyrochronology. This was done using the open-source python package \texttt{stardate} \citep{2019AJ....158..173A}. \texttt{stardate} uses a star's astrometric, photometric, spectroscopic, and rotational information, along with their associated uncertainties, to provide an age constraint through mcmc sampling of the \texttt{MIST} evolutionary models \citep{2016ApJ...823..102C, 2016ApJS..222....8D}. For the input parameters of BD+24 we used the astrometry and photometry from Gaia DR3, as well as photometry from 2MASS \citep{2MASS} and the V magnitude from the Hippparcos input catalog \citep{1993BICDS..43....5T}. \cite{2018AJ....155...39O} report a rotational period for BD+24 of P$_{rot}=30.42$ d which we also used as an input for \texttt{stardate}. For the spectroscopic parameters, we used the $\log~g$, [Fe/H], and $T_{\rm eff}$ derived in Section \ref{ses:abundances}. This provides a tight age constraint of 3.3$^{+0.7}_{-0.5}$ Gyr. All other age diagnostics are consistent with this value, so we adopt the \texttt{stardate} derived age for BD+24. 

\section{CW2106 Characterization} \label{comp_characterization}

\subsection{CW2106 Spectral Type} \label{sec:sp_types}
%To assign a spectral type to CW2106, we first followed the approach of \cite{Kirkpatrick2010}. The JWST/NIRSpec spectrum of CW2106 was normalized to 1.28 $\mu$m and the spectral standards from \cite{Kirkpatrick2010} were over-plotted and normalized in the same manner. The J-bands of both CW2106 and the standards were then visually compared. Figure \ref{fig:sptype_all} shows the NIRSpec data along with the L9 and T0 standards, the two best fits visually. Both the L9 and T0 fit the J-band of CW2106 equally well, making it difficult to discern between the two. Looking at the overall continuum, the L9 provides the closest fit to CW2106, although the shape of the H-band of CW2106 appears closer to that of the T0 standard.

To assign a spectral type to CW2106, we first followed the approach of \cite{Kirkpatrick2010}. The JWST/NIRSpec spectrum of CW2106 was normalized to 1.28 $\mu$m and the spectral standards from \cite{Kirkpatrick2010} were over-plotted and normalized in the same manner. The J-bands of both CW2106 and the standards were then visually compared. We found that both the L9 and T0 spectral standards fit the J-band of CW2106 equally well, making it difficult to discern between the two. Looking at the overall continuum, while the L9 provided the closest fit to CW2106 in flux, the spectral morphology of CW2106 appeared to more closely resemble that of the T0 standard.

%\begin{figure}
%    \centering
%    \includegraphics[width=1.0\linewidth]{sptyping.jpeg}
%    \caption{\textbf{Top:} The JWST/NIRSpec spectrum of CW2106, over-plotted with the L9 (solid blue line) and T0 (solid purple line) spectral standards from \cite{Kirkpatrick2010}. All spectra have been normalized to 1.28 $\mu$m. \textbf{Bottom:} The signal-to-noise (SNR) of the JWST/NIRSpec spectrum.}
%    \label{fig:sptype_all}
%\end{figure}

\begin{figure*}
    \centering
    \includegraphics[width=1.0\textwidth]{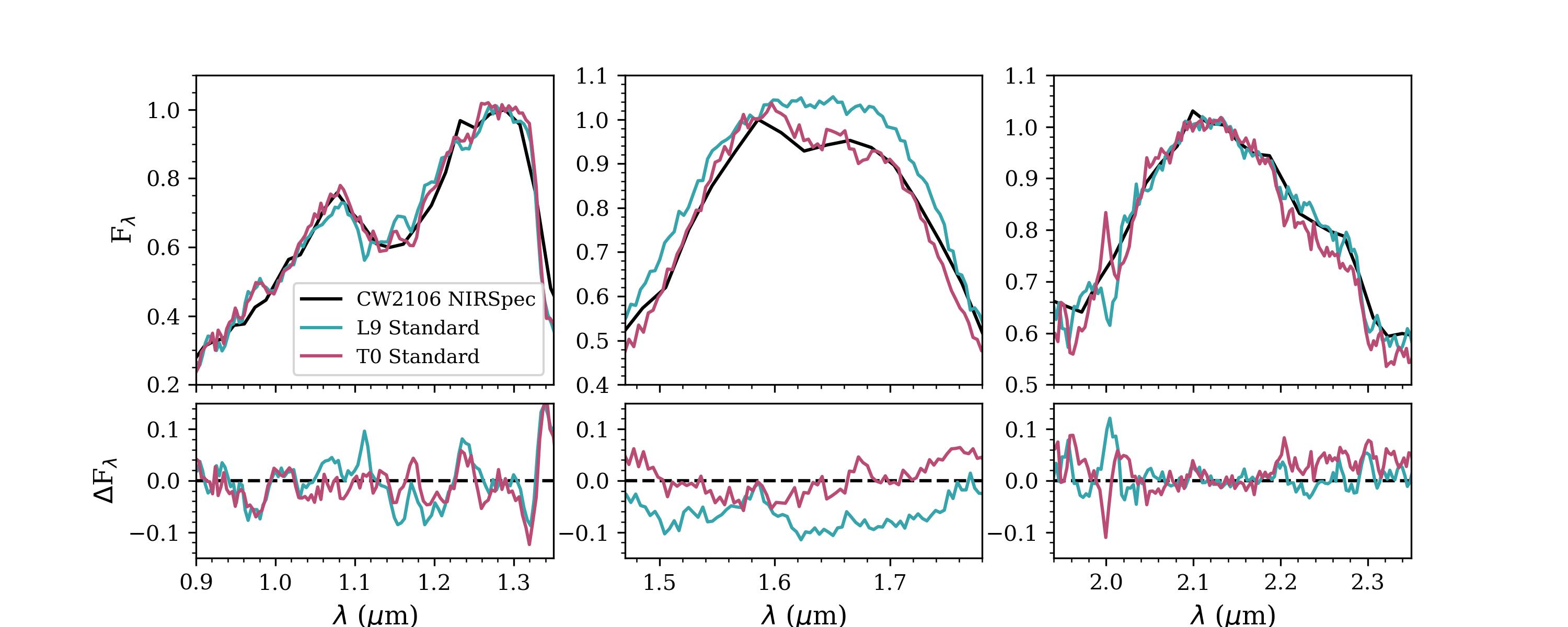}
    \caption{\textbf{Top Row:} $0.9-2.35 ~\mu$m near-IR spectrum of CW2106 (black) along with the L9 (blue) and T0 (purple) standards, split into the J- (left panel), H- (middle panel), and K-band (right panel). Each spectrum was normalized to the peak of each respective band. \textbf{Bottom Row:} Residuals between CW2106 and the L9 (blue) and T0 (purple) standards for the J-, H-, and K-bands (left, middle, right panels respectively).}
    \label{fig:sptype_bandByband}
\end{figure*}

To better help compare the spectral morphology of CW2106, we next compared each of the J-, H-, and K-bands separately, rather than the shape of the continuum. To do this, we cut each spectrum into the three individual bands, normalizing each to their respective peaks in flux. Figure \ref{fig:sptype_bandByband} shows the normalized J-, H-, and K-bands of CW2106 compared to the L9 and T0 standards. While the K-band shape is better fit by the L9, the overall shapes of the J- and H-Bands of CW2106 are more similar to those of the T0 standard. We therefore assign CW2106 a spectral type of T0$\pm$1.

\subsection{CW2106 SED \& Fundamental Parameters} \label{sec:SED}
We used the open-source code SEDkit \citep{2020ascl.soft11014F, sedkit} in order to obtain the flux calibrated spectral energy distribution (SED) for CW2016+2507. In this section, we summarize the methodology employed by SEDkit, as well as present the SED and derived properties of CW2106. 

\subsubsection{Construction of SED}
In order to create an absolute flux calibrated SED and empirically derive bolometric luminosity ($L_{\rm bol}$), SEDkit follows the approach presented in \cite{2015ApJ...810..158F}. The first step is to provide SEDkit with any available spectra and their associated errors. For these we used the JWST NIRSpec/PRISM and MIRI/LRS spectra described in Section \ref{sec:obs-jwst}. In order to ensure the best quality data, we removed the detector edge portions of the NIRSpec and MIRI spectra (corresponding to SNR$<$5), resulting in continuous spectral coverage between 0.83--12.5 $\mu$m. A composite spectrum is then created using the average flux in their regions of overlap, which for the NIRSpec/PRISM and MIRI/LRS spectra is between 4.9 -- 5.5 $\mu$m. We also provided SEDkit with literature photometry (Table \ref{tab:sys_prop}) for CW2106. SEDkit then linearly interpolates from 0 $\mu$m to the shortest wavelength data point and appends a Rayleigh-Jeans tail from the longest wavelength data point out to 1000 $\mu$m. 

\begin{figure*}
    \centering
    \includegraphics[width=1\textwidth]{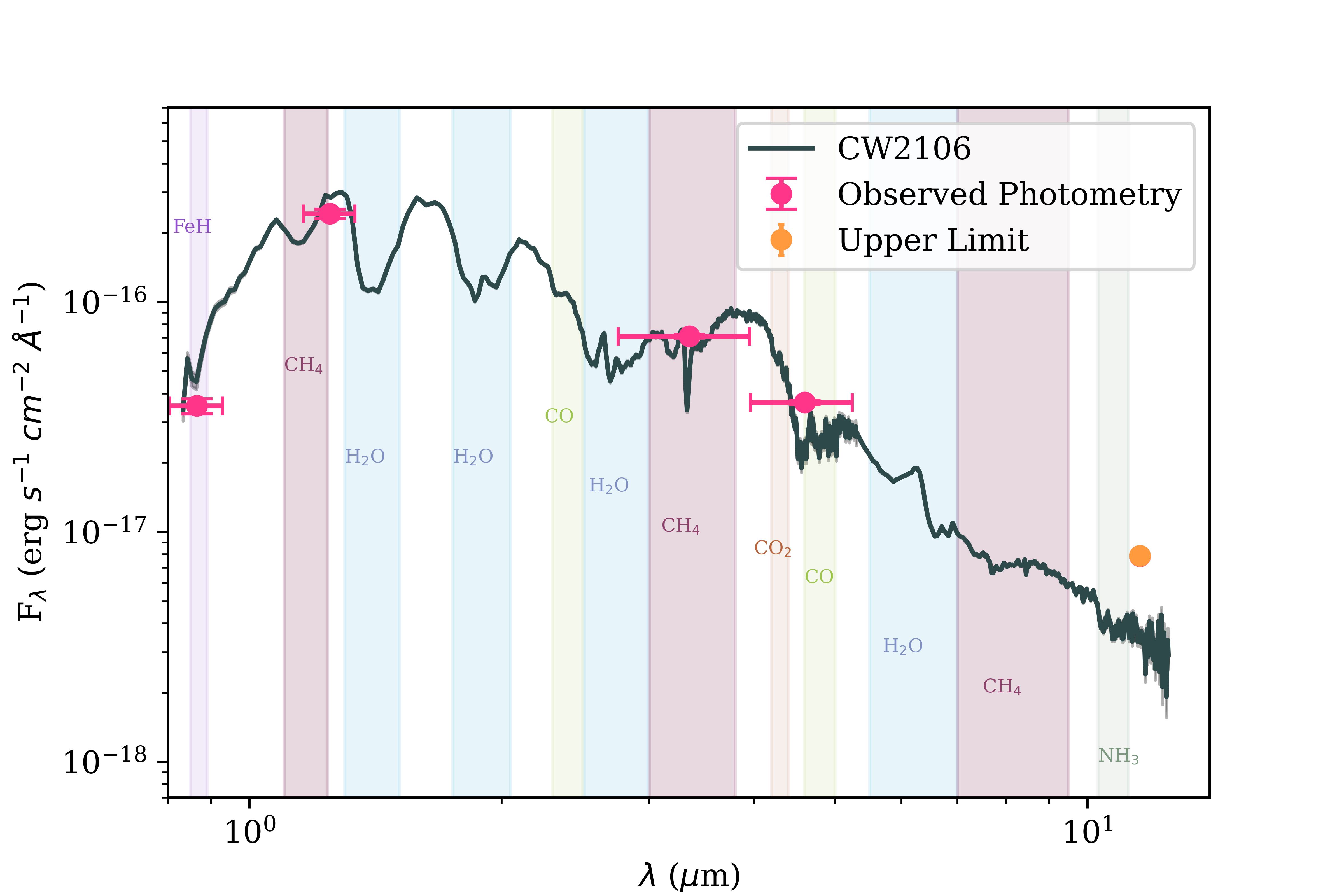}
    \caption{The Full (97.3\%) JWST SED of CW2106 shown in black. The observed photometry for CW2106 (listed in Table \ref{tab:sys_prop}) of CW2106 are overplotted as pink points, with the associated flux errors and bandpass widths. Regions of broad molecular absorptions are highlighted and labeled.}
    \label{fig:sed_comparison}
\end{figure*}

Figure \ref{fig:sed_comparison} shows the final, flux calibrated SED for CW2106. The near-simultaneous, broad spectral coverage of CW2106 provides an in depth view of its atmosphere in a way not possible without JWST. For example, we can take a look at a similar object: SDSS J120747.17+024424.8 (J1207), a field dwarf of spectral type T0 \citep{Burgasser_etal2006}. \cite{Stephens_etal2009} presented the SED of J1207, which at the time was one of the broadest wavelength coverage SEDs for an L/T transition object. However, to achieve a similar spectral coverage to CW2106, J1207 required the use of multiple ground- and space-based facilities, an observationally expensive endeavor, while still leaving several large gaps of wavelength. It should be pointed out that pre-JWST, assembling an SED as complete as J1207 was only possible for the brightest objects, leaving the majority of brown dwarf SEDs much more incomplete. The observations of J1207 were carried out over several years (SDSS - July 2001; UKIRT/CGS4 Z - Jan 2004; UKIRT/CGS4 J - July 2002; UKIRT/CGS4 H \& K - Jan 2001; Gemini/NIRI - April, May 2005; \textit{Spitzer}/IRS - June 2005) and, as brown dwarf atmospheres are dynamic, possibly cover different atmospheric conditions. The SED for CW2106 from JWST is not plagued by these challenge, allowing for its atmosphere to be investigated under the same conditions and with no gaps in spectral coverage, highlighting just one way JWST will be able to help unlock the mysteries of these dynamic, enigmatic worlds.

\subsubsection{Derivation of Fundamental Parameters} \label{sec:fund-params}

With the full flux calibrated SED created, SEDkit is then able to empirically derive the bolometric luminosity ($L_{\rm bol}$) for CW2106. This is done using the equation: 
\begin{equation}
    L_{bol} = 4\pi d^2 \int^{1000\mu m}_{0\mu m} F_{\lambda}d\lambda
\end{equation}
where $d$ is the distance and $F_{\lambda}$ is the absolute flux calibrated SED. For the distance of CW2106, we adopted the Gaia DR3 distance of its host star of $34.11\pm0.02$ pc \citep{GaiaDR3}.  

The derived $L_{\rm bol}$ is then used by SEDkit along with evolutionary models and a provided age range to estimate the object's mass, radius, and $\log~g$. The models chosen for this analysis were the evolutionary tracks of \cite{2008ApJ...689.1327S} (SM08), which model the L/T transition through use of a ``hybrid'' cloudy atmosphere prescription that linearly interpolates from a 1400 K cloudy atmosphere to a cloudless atmosphere at 1200 K. Model independent determinations of mass and luminosity for individual components of brown dwarf binaries have shown SM08 to be in excellent agreement with the data for objects within the L/T transition \citep[i.e.][]{Dupuy_Liu2017}. With a spectral type of T0$\pm$1 (Section \ref{sec:sp_types}), CW2106 falls right in the middle of the L/T transition, and should therefore be best modeled by SM08. 

For the age of CW2106, we are once again able to turn to the host star and adopt the age for BD+24 which was estimated in Section \ref{sec:age}. The effective temperature ($T_{eff}$) of CW2106 was then calculated using the Stefan-Boltzmann Law:
\begin{equation} \label{eq:2}
    T_{eff} = (\frac{L_{bol}}{4 \pi R^2 \sigma_{SB}})^{1/4}
\end{equation}
where $R$ is the estimated radius of CW2106, and $\sigma_{SB}$ is the Stefan-Boltzmann constant. 

Uncertainties were estimated using the spread of values derived repeating this analysis with the upper and lower bounds of the distance, age, and flux. A summary of all derived and estimated parameters for CW2106 is shown in Table \ref{tab:sys_prop}.

\subsubsection{Comparison to Literature} \label{sec:param_comp_lit}
With the fundamental parameters of CW2106 now in hand, it is possible to see how it compares with other sub-stellar objects. The spectral type of CW2106 was found in Section \ref{sec:sp_types} to be closest to T0$\pm$1; previous studies have found field T0 dwarfs to have a $T_{\rm eff}$ of 1247$\pm$113 K \citep{2015ApJ...810..158F}  or 1190$\pm$175 \citep{2023ApJ...959...63S}. The $T_{\rm eff}$ for CW2106 in Section \ref{sec:fund-params} of 1213$\pm21$ K is consistent with these estimates, placing CW2106 as typical for its spectral type.

Using the continuous, broad wavelength coverage from JWST we are able to constrain our $L_{\rm bol}$ measurement to $<$0.5\% uncertainty, adding it to the growing list of brown dwarfs observed by JWST with precise $L_{\rm bol}$ \citep[e.g.,][]{Beiler_etal2024B}. However, even with extremely precise $L_{\rm bol}$ measurements, \cite{Beiler_etal2024B} find that their uncertainty in $T_{\rm eff}$ is limited to 2.5\%, a result of poorly constrained ages or radii. The benchmark nature of CW2106 means that we are able to adopt a precise distance and age (therefore radius) via BD+24, overcoming this obstacle, and thus we are able to achieve an uncertainty in $T_{\rm eff}$ of $\sim1.7\%$, a significant improvement over the typical error in $T_{\rm eff}$ of field L/T transition objects from previous works ($\sim$9\%; \citealt{2015ApJ...810..158F}).

The mass of CW2106 was estimated by \citep{Rothermich24} to be $\sim$36 M$_{Jup}$. This estimate was made using the phototype of CW2106 in \cite{Rothermich24} ($\sim$T2) and the average mass per spectral type bin relation derived in \cite{Dupuy_Liu2017} using objects with dynamical mass measurements, which placed CW2106 within the T2-T5.5 bin with a mass of 36$\pm$9 M$_{Jup}$. In Section \ref{sec:sp_types}, we found that the spectral type of CW2106 was T0$\pm$1, which would actually place it within the \cite{Dupuy_Liu2017} spectral type bin of L8-T1.5, which has a mass range of 34--62 M$_{Jup}$. We derive for CW2106 a mass range of 50--62 M$_{Jup}$, consistent with the value from this new spectral type bin.

\begin{figure}
    \centering
    \includegraphics[width=1.0\linewidth]{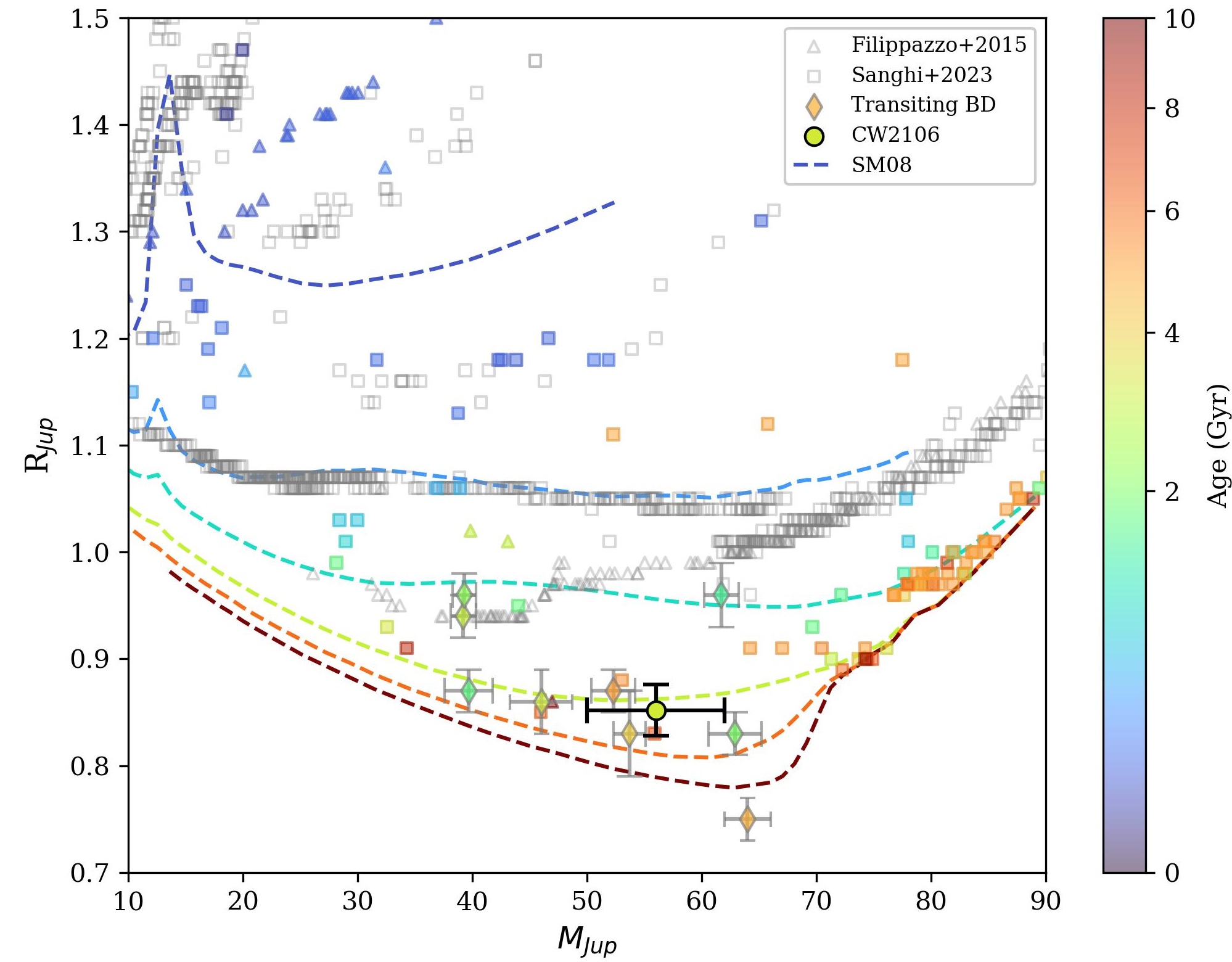}
    \caption{M$_{\rm Jup}$ vs R$_{\rm Jup}$ for the sample from \cite{2015ApJ...810..158F} (F15, triangles), the sample from \cite{2023ApJ...959...63S} (S23, squares), and CW2106 (circle). Also shown as diamonds are the transiting brown dwarfs WASP-128b \citep{Hodzic2018}, CWW 89Ab \citep{Nowak2017}, KOI-205b \citep{Diaz2013}, TOI-1406b \citep{Carmichael2020}, EPIC 212036875b \citep{Carmichael2019, Persson2019}, TOI-852b \citep{Carmichael2021}, WASP-30b \citep{Anderson2011}, LHS 6343C \citep{Johnson2011}, and TOI-569b \citep{Carmichael2020}, using the updated values presented by \cite{Carmichael2023}. Overplotted as dashed lines are the evolutionary models of (\citealt{2008ApJ...689.1327S}, SM08). All points and lines are colored by their assigned age as indicated by the colorbar, or if no age constraint is present, points are left as empty grey outlines. } 
    \label{fig:Mass_v_Rad}
\end{figure}

Figure \ref{fig:Mass_v_Rad} plots mass vs. radius for CW2106 and the F15 \& S23 samples, colored based on assigned age, along with the SM08 evolutionary models. Also shown in Figure \ref{fig:Mass_v_Rad} are transiting brown dwarfs with similar masses and ages to CW2106, who's properties have been independently determined. This comparison shows that while CW2106 has a smaller radius than the bulk F15 \& S23 samples, this is largely due to the lack of constrained ages for these objects. When only looking at the F15 and S23 objects with ages, CW2106 becomes less of an outlier, falling in line with the expected radii for objects $>$1 Gyr as indicated by the evolutionary model tracks. Figure \ref{fig:Mass_v_Rad} also shows the derived radius of CW2106 is in good agreement with the independently measured radii from the transiting brown dwarfs.

\section{Spectral Analysis} \label{sec:spec_analysis}
In this section we present an analysis of the spectral features of CW2106 using the Magellan/FIRE, NIRSpec/PRISM, and MIRI/LRS spectra. We divide our discussion into three main sections: \textit{i}) the 0.8-2.5 $\mu$m portion of the near-IR (the ``classical'' near-IR), \textit{ii}) the 2.5-5 $\mu$m portion of the near-IR (uniquely obtained from state-of-the-art space based observatories, referred to from this point as the near/mid-IR), and \textit{iii}) the mid-IR from 5-12.5 $\mu$m.

\subsection{0.8-2.5 $\mu$m}
\begin{figure*}
    \centering
    \includegraphics[width=1.0\textwidth]{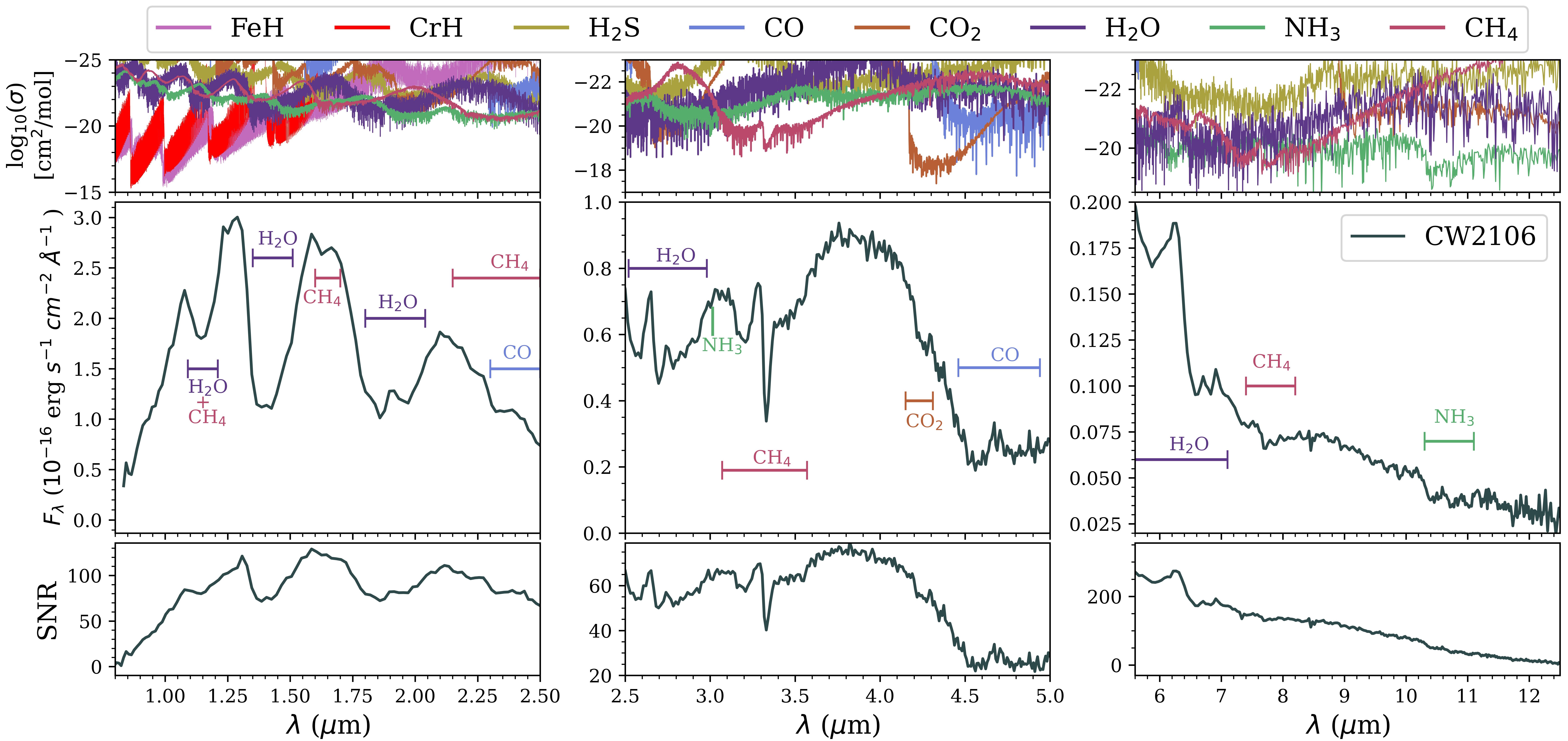}
    \caption{\textbf{Left:} Absorption cross sections for several gasses common in brown dwarf atmospheres taken from \cite{Hood_2024} and references within (top panel). Cross sections are shown for $T_{\rm eff}$=1200K and a pressure of 1 bar. The $0.8-2.5 ~\mu \rm m$ NIR region of the JWST/NIRSpec spectrum for CW2106 shown in the middle panel in black. Broad molecular absorption features are labeled in the same color as their respective cross sections. The corresponding SNR of CW2106 is shown in the bottom panel. \textbf{Middle:} Same as the left, but for $2.5-5 ~\mu \rm m$ near/mid-IR. \textbf{Right:} Same information shown for the mid-IR ($5-12.5) ~\mu \rm m$}.
    \label{fig:spec_analysis1}
\end{figure*}

Figure \ref{fig:spec_analysis1} shows the 0.8-2.5 $\mu$m near-IR spectrum of CW2106 from NIRSpec/PRISM, along with the absorption cross sections for the major absorbers expected (shown for T$_{eff}$=1200\,K at 1 bar) to exist within its atmosphere. Prominent water (H$_2$O) and methane (CH$_4$) absorption features dominate this region of the spectrum, sculpting the continuum into four distinct peaks at 1.08, 1.27, 1.59, and 2.07\,$\mu$m as is typical for this spectral type \citep{Burgasser_etal2002, Kirkpatrick_etal2005}. Also visible is the carbon monoxide (CO) feature at 2.3\,$\mu$m, and a large feature at 0.87\,$\mu$m due to a combination of chromium hydride (CrH) and iron hydride (FeH).  

Another weaker CrH+FeH feature at 0.97 $\mu$m can be seen in the FIRE/Echelle spectrum (Figure \ref{fig:fire_spectrum}). Figure \ref{fig:fire_spectrum} also labels the K I doublets at $\sim$1.17 and 1.25 $\mu$m. While they are visible in the FIRE/Echelle spectrum, the features are very weak. The spectrum is visibly very noisy, so the weak lines may be due to the low SNR of the data more-so than an inherent property of CW2106. K I strength has been found to be correlated with surface gravity for late-M and L-dwarfs \citep[e.g.,][]{McGovern2004}, where the strengths of the lines decrease with decreasing surface gravity. However, \cite{Suarez_etal2021} found that K I strength is no longer a reliable indicator of surface gravity once into the early-T spectral types ($\sim$T2-T3). If the observed K I strengths of CW2106 are reliable, however, its weak lines plus high surface gravity ($\log~g$$\sim5.28$ dex) would indicate that this lack of correlation with surface gravity could possibly begin as early as spectral type T0. Future higher spectral resolution observations of CW2106 will be able to help determine the strengths of its K I doublets, constraining their behavior during a critical transition point in brown dwarf evolution. 

\subsection{2.5-5 $\mu$m}

The 2.5-5 $\mu$m portion of CW2106's NIRSpec/PRISM spectrum is shown in Figure \ref{fig:spec_analysis1} with the absorption cross sections for the expected major absorbers. Several striking features are observed including broad H$_2$O at $\sim$2.7 $\mu$m, the 3.3 $\mu$m fundamental CH$_4$ band, carbon dioxide (CO$_2$) at 4.2 $\mu$m, and the 4.6 $\mu$m fundamental CO band. 

We also tentatively identify the 3 $\mu$m $\nu _3$ NH$_3$ band first identified by \cite{Beiler_etal2023} and confirmed by \cite{Beiler_etal2024B}, which analyzed the spectra of 23 late-T and Y dwarfs (ranging from $T_{\rm eff}\sim1,000$ K to 350 K). While the proposed feature in CW2106 lines up with the increase of NH$_3$ opacity from the absorption cross sections in Figure \ref{fig:spec_analysis1} and agrees with what is seen in \cite{Beiler_etal2024B}, we cannot confirm this detection due to its low strength. \cite{Beiler_etal2024B} found the depth of the $\nu _3$ NH$_3$ band increased with decreasing $T_{\rm eff}$ in the majority of their sample, so if confirmed in CW2106, its warmer $T_{\rm eff}$ (1213$\pm21$ K) and shallower absorption depth would agree with this trend. Future observations at a higher spectral resolution may help to confirm this detection. 

%\begin{figure}
%    \centering
%    \includegraphics[width=1.0\linewidth]%{Spectral_analysis_2.jpeg}
%    \caption{Similar to Figure \ref{fig:spec_analysis1}, but over the 2.5--5 $\mu$m region.}
%    \label{fig:spec_analysis1}
%\end{figure}

\subsection{5-12 $\mu$m} \label{miri-spec-analysis}
Figure \ref{fig:spec_analysis1} shows the 5--12 $\mu$m MIRI/LRS spectrum for CW2106. The \textit{Spitzer} IRS sample from \cite{Suarez_etal2022} shows that for objects of a similar spectral type to CW2106 (T0$\pm$1) this region is dominated mainly by prominent absorption features of H$_2$O and CH$_4$. These features are clearly seen in the MIRI/LRS spectrum of CW2106, and have been labeled in Figure \ref{fig:spec_analysis1}. Also visible in the MIRI/LRS spectrum is the 10.5 $\mu$m NH$_3$ feature. \cite{Suarez_etal2022} found that NH$_{3}$ became detectable as early as spectral type T1.5 in their IRS sample, however CW2106 is a whole subtype earlier. One potential explanation for the lack of NH$_3$ at earlier spectral types in the IRS sample for spectral types $<$T1.5 is the low SNR of the \textit{Spitzer} spectra. The median SNR of the spectra in \cite{Suarez_etal2022} is $\sim$20 at 6 $\mu$m and $\sim$10 at 12 $\mu$m. While the MIRI/LRS spectrum also has an SNR of $\sim$10 at 12 $\mu$m, its SNR at 6 $\mu$m is closer to $\sim$250. This increase in SNR may explain the detection of NH$_3$ in CW2106, and future JWST observations of similar type objects may be able to confirm whether or not the 10.5 $\mu$m NH$_3$ feature is common at this spectral type. 

We find no evidence of the $\sim$8--11 $\mu$m silicate absorption feature for CW2106 in Figure \ref{fig:spec_analysis1}. \cite{Suarez_etal2022} found using the \textit{Spitzer} IRS sample that the onset of the silicate feature in field dwarfs is around spectral type L2, peaks around L4-L6, and is last detected at L8. In young brown dwarfs, clouds have been inferred as late as T2.5 with retrieval modeling \citep[e.g.,][]{Vos_etal2023}, likely due to lower surface gravity allowing clouds to form higher up in the atmosphere. The strength of the silicate feature has also been found to strongly depend on the observed inclination angle \citep{Suarez_2023}, with equator-on objects displaying a stronger silicate absorption than objects viewed pole-on. At spectral type T0$\pm$1 with an age of 3.3$^{+0.7}_{-0.5}$ Gyrs, its not surprising CW2106 lacks a noticeable silicate feature, especially if it is at a more pole-on inclination, following the trend observed in older field dwarfs.

%\begin{figure}
%    \centering
%    \includegraphics[width=1.0\linewidth]{Spectral_analysis_3.jpeg}
%    \caption{Same as in Figure \ref{fig:spec_analysis1} but for the MIRI/LRS spectrum of CW2106 (in black).}
%    \label{fig:spec_analysis1}
%\end{figure}

\subsection{Spectral Indices} \label{sec:indices}
We calculated the CH$_4$, NH$_3$, and H$_2$O indices as defined by \cite{2006ApJ...648..614C} for CW2106 using its MIRI LRS spectrum, allowing for a quantitative comparison of its spectral features discussed in Section \ref{miri-spec-analysis} with those observed in the broader population. These indices focus on the 6.25 $\mu$m H$_2$O bands, the 8.5 $\mu$m CH$_4$ band, and the 10.5 NH$_3$ band. 

For the calculations, we followed the approach of \cite{Suarez_etal2022}, which modifies the CH$_4$ index to focus on the 7.65 $\mu$m region rather than 8.5 $\mu$m, as well as defining an index for the silicate absorption feature centered around 9 $\mu$m. These indices are defined as:
\begin{equation}
    \textnormal{H$_2$O Index}\ = \frac{F_{6.25}}{0.562F_{5.8} + 0.474F_{6.75}}
\end{equation}
\begin{equation}
    \textnormal{CH$_4$ Index} = \frac{F_{10.0}}{F_{7.65}}
\end{equation}
\begin{equation}
    \textnormal{NH$_3$ Index} = \frac{F_{10.0}}{F_{10.8}}
\end{equation}
\begin{equation}
    \textnormal{Silicate Index} = \frac{C_{9.3}}{F_{9.3}}
\end{equation}
where $F_{\lambda}$ is the average flux around the specified wavelength, using a window of 0.3 $\mu$m for the H$_2$O index and 0.6 $\mu$m for CH$_4$, NH$_3$, and the silicate index. $C_{\lambda}$ is defined in \cite{Suarez_2023} as the average flux of a continuum assuming an exponential interpolation between 7.5 and 13.5 $\mu$m. However, we chose to use 11.5 $\mu$m vs 13.5 $\mu$m as the red end of the interpolation, as the SNR of the MIRI/LRS data drops significantly beyond $\sim12 \mu$m. Errors were computed using the errors of the the average flux in each window. The indices are designed such that values $>1$ indicate the presence of the molecule as long as the molecule feature region is not affect by other spectral features , with larger values indicating a stronger absorption feature.  

\begin{figure*}
    \centering
    \includegraphics[width=1.0\textwidth]{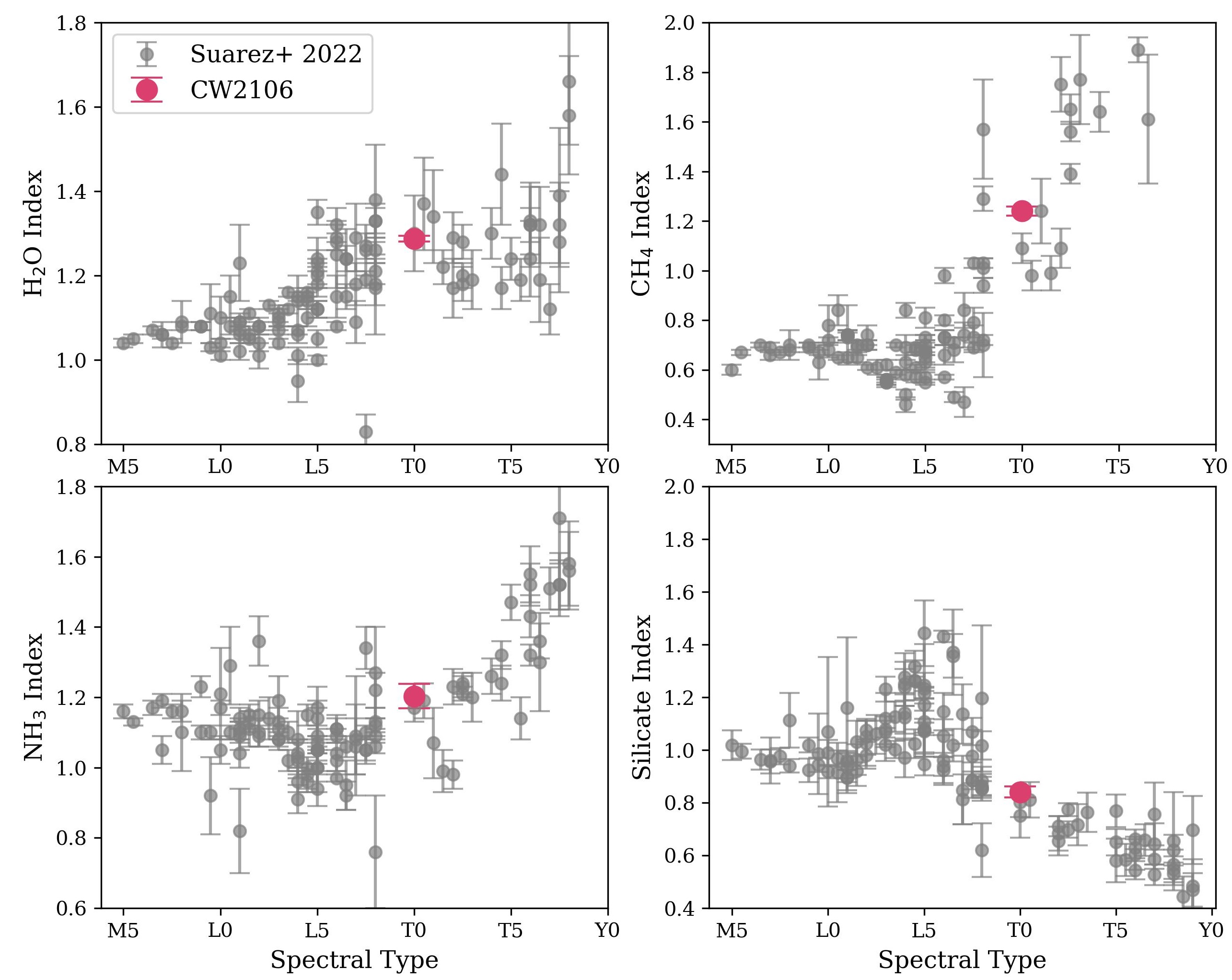}
    \caption{Molecular indices as defined in Section \ref{sec:indices} vs spectral type. CW2106 is shown as a red circle, with the \textit{SPITZER} IRS sample from \cite{Suarez_etal2022} shown as grey circles.} 
    \label{fig:spt_indices}
\end{figure*}

Figure \ref{fig:spt_indices} shows the H$_2$O, CH$_4$, NH$_3$, and silicate indices vs spectral type for CW2106 compared to the \textit{Spitzer} IRS sample from \cite{Suarez_etal2022}. The higher SNR provided by JWST allows for more precise molecular indices than what was possible for most \textit{Spitzer} observations. This is noticeable in Figure \ref{fig:spt_indices}, with perceptibly smaller error bars for CW2106 over the IRS sample, highlighting the power of JWST to better characterize and constrain the properties of brown dwarfs.   

Comparing CW2106 with objects of a similar spectral type, the H$_2$O index appears near normal for its spectral type. The CH$_4$ index for CW2106 is slightly elevated, although the spread of CH$_4$ indices of L/T transition objects is quite large. CW2106's NH$_3$ index also appears to be slightly higher than objects of a similar spectral type, though agrees within the error bars. Consistent with similar type objects from the \textit{Spitzer} IRS sample, the silicate index for CW2106 indicates the lack of a silicate absorption feature. This is in agreement with the lack of visual evidence for silicates in the mid-IR spectrum of C2106 in Section \ref{miri-spec-analysis}.

\section{Comparison with Forward Models} \label{sec:modeling}

In this section, we compare the full 0.8--12.5 $\mu$m SED of CW2106 with a series of self-consistent atmospheric grid models encompassing both cloudy/cloudless atmospheres and equilibrium/disequilibrium chemistry. In order to do this, we used the open-source python code \texttt{SEDA}\footnote{\url{https://seda.readthedocs.io/en/latest/index.html}} (\citealt{Suarez_etal2021}, Suarez et al. in prep), which provides $\chi^2$ minimization and nested sampling frameworks for comparing observed data to multiple different atmospheric model grids. We performed a $\chi^2$ minimization over all of the model grids chosen (described below) in order to find the best fits. 

In addition to finding the best model fits to the full SED, we also investigated the best fitting models over 3 different wavelength ranges using those defined in Section \ref{sec:spec_analysis}: 0.8--2.5 $\mu$m (the ``classical'' near-IR), The 2.5--5 $\mu$m near/mid-IR, and 5--12 $\mu$m (the mid-IR). The near- and mid-IR originate in different layers of the atmosphere; lower pressures ($\lesssim$1 bar, \citealt{Burningham_etal2021}) are probed by the mid-IR while the near-IR probes the hotter, deeper layers ($\sim$1-20 bar, \citealt{Karalidi_etal2021, McCarthy_etal2024}). Exploring which models best fit these different wavelengths therefore yields insight into the potential conditions present in different layers of the atmosphere.  
 
\subsection{Models Used}
To perform the $\chi^2$ minimization fits, we used three separate grid models covering different chemistry regimes and atmospheric parameters. Below we provide a brief description of the models used, their assumptions, and the parameter space covered. For a more detailed discussion of a specific model grid, we direct the reader to that model's cited paper.  

\subsubsection{\texttt{ATMO2020}}
The \texttt{ATMO2020} atmospheric models \citep{Phillips_etal2020} are a set of three grids which assume solar metallicity and a cloudless atmosphere. Each grid includes a different regime of chemistry: \textit{i)} chemical equilibrium, \textit{ii)} non-equilibrium chemistry caused by weak vertical mixing, parameterized with by the eddy diffusion coefficient K$_{zz}$ ($log(K_{zz})=4$ dex), and \textit{iii)} non-equilibrium chemistry due to strong K$_{zz}$ ($log(K_{zz})=6$ dex). All three grids cover the range $T_{\rm eff}$=200--3000 K (with steps of 100 K and 50 K for T$_{eff}>600$ K and  T$_{eff}<600$ K respectively) and $\log~g$=2.5--5.5 dex (with steps of 0.5 dex).

\subsubsection{\texttt{Sonora Elf Owl}}

The \texttt{Sonora Elf Owl} atmospheric models were developed by \cite{Mukherjee_etal2024} as part of the Sonora family of models. This model assumes a cloudless atmosphere, disequilibrium chemistry due to vertical mixing, and takes into account the effects of changing metallicity and C/O ratio. The \texttt{Elf Owl} grid spans the parameter space $T_{\rm eff}$=275--2400 K, $\log~g$=3.25--5.5 dex, log(K$_{zz}$)=(2,4,7,8,9) dex, [M/H]=(-1.0, -0.5, 0.0, +0.5, +0.7, +1.0) dex, and C/O= (0.22, 0.458, 0.687, 1.12). The $\log~g$ varies by 0.25 dex and the $T_{\rm eff}$ varies by 25 K between 275--600 K, 50 K between 600--1000 K, and 100 K between 1000--2400 K.  

\subsubsection{\texttt{Sonora Diamondback}}

Another in the Sonora family of models, \texttt{Sonora Diamondback} \citep{Morley_etal2024} is a grid of atmospheric models under chemical equilibrium which include the effects of clouds at three different metallicities. Cloud species included are enstatite (MgSiO$_3$), forsterite (Mg$_2$SiO$_4$), aluminum oxide (Al$_2$O$_3$), and iron (Fe), with varying thickness for all species (controlled by the parameter $f_{sed}$). The \texttt{Diamondback} models span the parameter space $T_{\rm eff}$=900--2400 K ($\Delta$$T_{\rm eff}$=100 K), $\log~g$=3.5--5.5 dex ($\Delta$$\log~g$=0.5 dex), [M/H]=(-0.5, 0.0, +0.5) dex, and $f_{sed}$=(1, 2, 3, 4, 8, nc), where ``nc'' indicates a cloudless atmosphere. All models are assumed to have a solar C/O.

\subsection{Results}

\subsubsection{Full SED Fit} \label{sec:fullm fit}

\begin{figure*}
    \centering
    \includegraphics[width=1.0\textwidth]{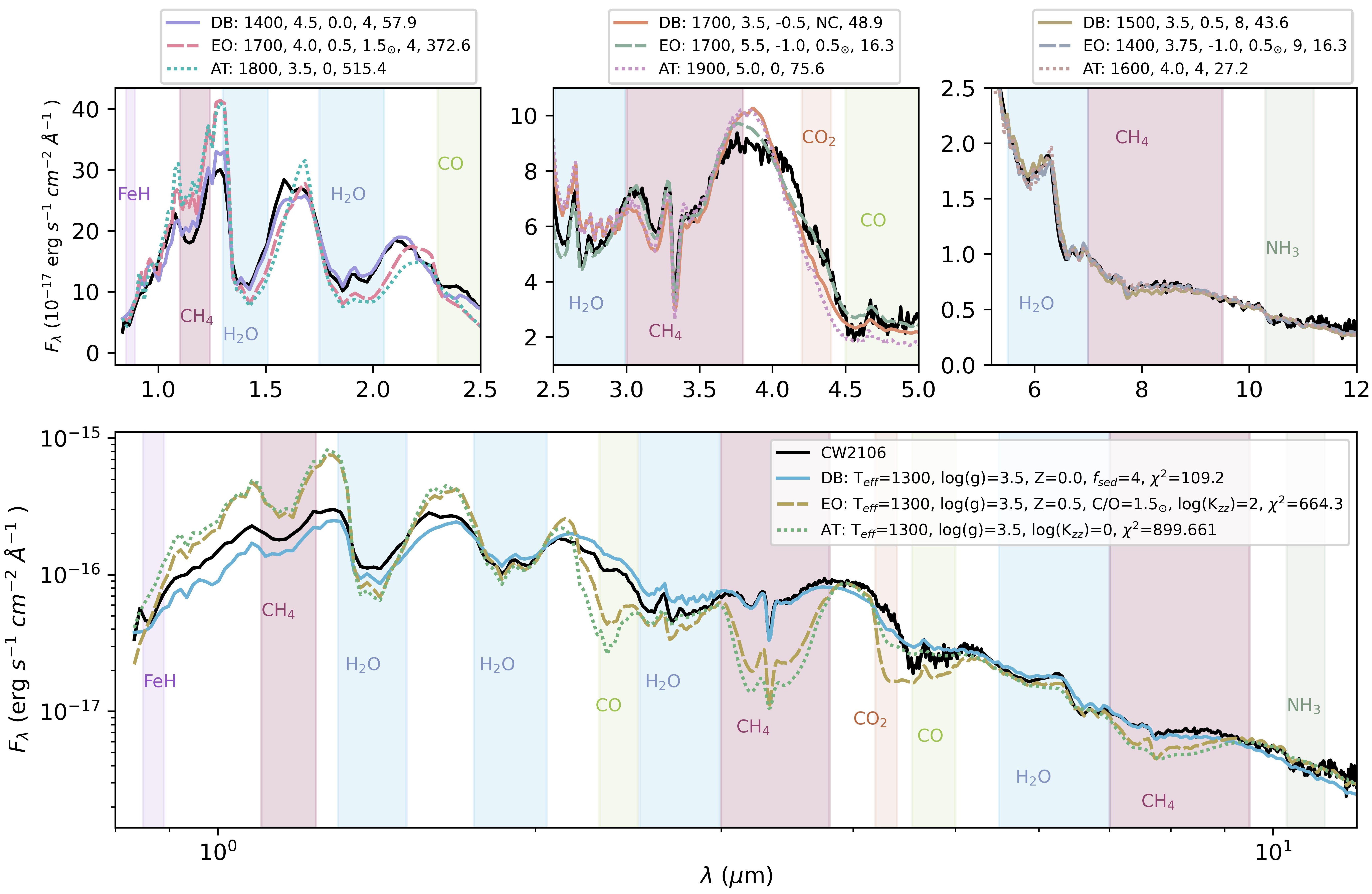}
    \caption{The SED of CW2106 (solid black line) plotted with the best fitting \texttt{Sonora Diamondback} (SD; colored solid line), \texttt{Sonora Elf Owl} (EO; colored dashed line), and \texttt{ATMO2020} (AT; colored dotted line) models. Regions of broad molecular absorptions are highlighted and labeled. The best fits to the 0.80--2.5 $\mu$m, 2.5--5 $\mu$m, and 5--12 $\mu$m regions are shown in the left, middle, and right panels of the top row, with the bottom panel showing the best fits to the full SED. Model parameters are given in the same order as in the bottom panel: For SD - [$T_{\rm eff}$, $\log~g$, [M/H] (Z), $f_{sed}$, $\chi^2$]; For EO - [$T_{\rm eff}$, $\log~g$, [M/H] (Z), C/O, log(K$_{zz}$, $\chi^2$]; For AT - [$T_{\rm eff}$, $\log~g$, log(K$_{zz}$), $\chi^2$].}
    \label{fig:models}
\end{figure*}

Figure \ref{fig:models} shows the best fit models from each of the model families used for each of the four wavelength ranges compared (0.83--2.5 $\mu$m, 2.5--5 $\mu$m, 5--12.5 $\mu$m, and 0.83--12.5 $\mu$m) along with the SED of CW2106. Over the full 0.8--12 $\mu$m SED (bottom panel of Figure \ref{fig:models}), the \texttt{Sonora Diamondback} cloudy model (solid line) offers a significantly better fit over the cloudless \texttt{ATMO2020} (dotted line) and \texttt{Elf Owl} (dashed line). The preferred \texttt{Diamondback} model has a $T_{\rm eff}$=1300 K, $\log~g$=3.5 dex, [M/H]=0.0 dex, and $f_{sed}$=4. While this may be the best fitting model out of all of those compared ($\chi^2$=109.2 vs. 664.3 and 889.6 for \texttt{Elf-Owl} and \texttt{ATMO2020} respectively), visual inspection of the comparison shown in the bottom panel of Figure \ref{fig:models} shows that it is still a relatively poor fit to the data, particularly in the near-IR. The J- and H-bands of the \texttt{Diamondback} model are underluminous when compared with the observed flux from CW2106, whereas the model is overluminous in the $\sim$2.2--3 $\mu$m region. The \texttt{Diamondback} model also fails to capture the shape of the 4 $\mu$m peak and the depth of the 4.6 $\mu$m CO band absorption. The mere existence of CO in most brown dwarf atmospheres indicates of a deviation from equilibrium molecular abundances \citep[e.g.,][]{1997ApJ...489L..87N,Saumon2006,Sorahana_2012}, so the lack of disequilibrium chemistry in the \texttt{Diamondback} models may be responsible for this mismatch.       

The parameters of the best fitting \texttt{Elf-Owl} and \texttt{ATMO2020} models are similar to those of the \texttt{Diamondback} model; the only difference being the metallicity of the \texttt{Elf-Owl} model as [M/H]=0.5 dex (vs the \texttt{Diamondback} metallicity of [M/H]=0.0 dex). Both the \texttt{Elf-Owl} and \texttt{ATMO2020} fits favor a low vertical mixing rate, with log(K$_{zz}$)=2 dex and 0 dex (i.e. chemical equilibrium) respectively. Visual inspection of the fits in Figure \ref{fig:models} (as well as their $\chi^2$ values) confirms that the \texttt{Elf-Owl} and \texttt{ATMO2020} models do not adequately fit the observed data for CW2106 to the same degree as the \texttt{Diamondback} model. While the \texttt{Elf-Owl} and \texttt{ATMO2020} model fits are less favored than those from \texttt{Diamondback}, all three models agree on a $T_{\rm eff}$ of 1300 K, not very far from the value derived in Section \ref{sec:fund-params} ($T_{\rm eff}$=1213 K). This highlights the importance of obtaining broad wavelength coverage for the estimation of $T_{\rm eff}$, as model fits to smaller wavelength regions (see Section \ref{sec:individual_fits}) show a wide variance of $T_{\rm eff}$ depending on the model and wavelength region choices.

The preference for the cloudy \texttt{Diamondback} model is suggestive for the need of clouds in the atmosphere of CW2106 to explain its SED. While CW2106 lacks the 9 $\mu$m silicate feature (Section \ref{sec:spec_analysis}), its placement on the CMD in Figure \ref{fig:cmd} shows its colors are more similar to those to the cloudy L-dwarfs than those of the cloudless T-dwarfs. Hence, the interpretation of a cloudy atmosphere for CW2106 from the \texttt{Diamondback} fit is in agreement with the observed data. 

\subsubsection{Fits by Region} \label{sec:individual_fits}

Looking at the best fits for each of the chosen regions in the top three panels of Figure \ref{fig:models} shows a large degree of variance in the obtained parameters, both from the full SED model fits as well as between each region. These inconsistencies suggest that relying on any one small wavelength range may lead to incorrect results, as those analyses would be missing crucial spectral features which lie outside of that range. Instead, the consistent parameters between each of the models when compared to the full SED of CW2106 point towards the need for broad wavelength coverage in order to confidently infer an objects fundamental parameters. While the parameters such as $T_{\rm eff}$ and $\log~g$ for the individual fits in each region may not be be consistent, by comparing which model grid (and their different atmospheric assumptions) better fits the spectral features of CW2106, it may be possible to draw rough inferences as to the possible conditions at different depths within its atmosphere. 

Figure \ref{fig:models} shows that the \texttt{Diamondback} model provides a significantly better fit ($\chi^2$=57.9) to the 0.83--2.5 $\mu$m range of CW2106 over the cloudless \texttt{Elf-Owl} ($\chi^2$=372.6) and \texttt{ATMO2020} ($\chi^2$=515.4) models. This is consistent with the fits to CW2106's full SED, and likewise points to the need for clouds to explain the morphology of CW2106's near-IR spectrum. The fits to the 2.5--5 $\mu$m wavelength range tell a different story. Here, the best fitting model is from the cloudless \texttt{Elf-Owl} grid ($\chi^2$=16.3), followed by the \texttt{Diamondback} model ($\chi^2$=48.9) but with no clouds. The 5--12.5 $\mu$m mid-IR portion of CW2106's SED is similarly fit better by the cloudless models. 

The preference for cloudless models with the mid-IR wavelengths (which probe atmospheric pressures $\lesssim$1 bar) but the cloudy model in the near-IR (which probes pressures $\sim$1--20 bar) suggests that clouds are only present in the deepest visible layers of CW2106's atmosphere. As an L/T transition object, this may be evidence of the cloud sedimentation process ongoing within CW2106's atmosphere, having not yet rained out below the photosphere.

\section{Silicate Clouds} \label{sec:clouds}

\subsection{Clouds: A Quick Overview} \label{cloud_overview}

Early observations of L type brown dwarfs in the near-IR found that their significantly reddened colors were best explained by ``dusty'' atmospheres which were host to both solid and liquid condensate species \citep[e.g.][]{kirkpatrick1999,Kirkpatrick_etal2000,Chabrier2000,Akerman2001}. These observations were in agreement with theoretical models which predicted that by the effective temperatures of mid-L dwarfs, species such as liquid iron and magnesium silicates (e.g. SiO$_2$, MgSiO$_3$, Mg$_2$SiO$_4$) have formed within the observable atmosphere \citep{2006asup.book....1L}. These clouds act as an opacity source which blocks some of the light from deep within the atmosphere (observed primarily in near-IR wavelengths), redistributing the energy to longer wavelengths, resulting in a reddened SED. 

While redder than normal near-IR colors of brown dwarfs served as an indirect proxy for ``cloudiness'', it wasn't until the launch of the Spitzer Space Telescope \citep{Spitzer} that the first direct observational evidence of clouds was possible. Through an analysis of the 5-14 $\mu$m spectra of M, L, and T dwarfs from Spitzer, \cite{2006ApJ...648..614C} were able to show for the first time a prominent absorption feature from $\sim9-11$ $\mu$m which was not predicted by atmospheric models at the time. This feature was speculated to be due to absorption from the Si-O stretching vibration mode of solid silicate grains, particularly small grains ($<2$ $\mu$m) high in the atmosphere.

\cite{Suarez_etal2022} re-analyzed the entire M5-T9 Spitzer spectral sample, investigating the observed 9 $\mu$m silicate feature as a function of spectral type. This analysis found that the first appearance of the silicate feature is around spectral type L2 ($T_{\rm eff}\approx2000$ K). The silicate feature then peaks in strength around L4-L6 ($T_{\rm eff}\approx1700-1500$ K), and disappears after spectral type L8 ($T_{\rm eff}\approx1300$ K). Below this temperature, the clouds have sedimented to deeper layers of the atmosphere at higher pressures than those probed by mid-IR spectra ($\lesssim1$ bar). 

While the silicate feature in spectral type T0 and later has yet to be identified with any significance, cloud effects can still be seen elsewhere in an object's SED, particularly in the near-IR which probes the deeper pressures in which the clouds now sit. In addition to those which show redder than normal SEDs, several L/T transition objects have been found to be variable in the near-IR \citep[e.g.][]{Radigan2014,Eriksson2019}, thought to be driven by ``patchy'' clouds \citep{Burgasser2002_lt}. 

\cite{Vos_etal2023} investigated two such objects: SIMP J01365662+0933473 (T2.5, $T_{\rm eff}=1150\pm70$ K) and 2MASS J21392676+0220226 (T1.5, $T_{\rm eff}=1040\pm60$ K). Though both objects lack the 9-11 $\mu$m silicate feature, they are highly variable in the near-IR. \cite{Vos_etal2023} modeled each object using the \texttt{BREWSTER} Retrieval framework \citep[e.g.][]{Burningham_etal2017} to constrain both their cloud species, as well as their locations. They find that the data is best described by a patchy layer of Mg$_2$SiO$_4$ (located at 1.3-1.7 bar) above a deep iron cloud deck. Both objects investigated by \cite{Vos_etal2023} are young ($\sim200$ Myr), and thus expected to be cloudier than their field aged counterparts \citep{Suarez_2023}. However, these objects are excellent case studies in how the presence of clouds can be inferred from observed spectra, even without the detection of the 9 $\mu$m silicate feature.

\subsection{Is CW2106 Cloudy?} \label{sec:2106_is_cloudy}

The MIRI/LRS spectrum of CW2106 shows a clear lack of the 9 $\mu$m silicate absorption feature, indicating that clouds (if present) are not significantly impacting the observable photosphere. However, the model fits performed in Section \ref{sec:modeling} showed that the cloudy \texttt{Sonora Diamondback} models best fit the observed SED of CW2106 over the cloud-free \texttt{Sonora Elf-Owl} and \texttt{ATMO2020} models. CW2106 is also slightly redder than a typical T0, with J-W2=3.16$\pm$0.05 compared to J-W2=2.82$\pm$0.01 \citep{Kirkpatrick_etal2021}. These combined strongly suggest that CW2106 does in fact have silicate clouds in its observable photosphere, causing shallower gas absorption features and a slightly reddened SED. So why then does CW2106 lack a silicate feature?

One possible factor in why CW2106 may lack a silicate feature is its viewing angle. \cite{Suarez2023_viewingAngle} investigated the strength of silicate features for brown dwarfs with known inclination angles. They found that objects viewed equator-on had stronger silicate features than those viewed closer to the poles (i.e. the equatorial latitudes of brown dwarfs are cloudier). While no inclination measurements exist for CW2106 at this time, if CW2106 is viewed more pole on, this could help explain why no silicate feature is observed. However, this would not explain it's slightly reddened SED, as objects viewed pole on tend to have bluer than average near-IR colors \citep{Suarez2023_viewingAngle, Vos2017}, nor would it explain the preference for cloudy models to explain the observed data.

Instead, the lack of a 9 $\mu$m silicate feature may be due to the evolutionary stage of CW2106. As a T0 spectral type, CW2106 sits directly in the middle of the L/T transition - an evolutionary stage defined by the clearing of clouds from the observable photosphere. This process is driven by the ever cooling atmosphere allowing individual condensate particles to grow in size until they are no longer buoyant and ``rain out'' to the deeper layers of the atmosphere. As a result, the cloud layer would be observed to effectively sink to deeper and deeper layers of the atmosphere throughout the L/T transition, up until the point in which the clouds are too deep to be observed, thereby leaving the observable atmosphere ``cloud free''. Taking another look at the model fitting performed in Section \ref{sec:modeling}, when fitting the near- and mid-IR separately, while the near-IR was still best fit by the cloudy model, the cloudless models were preferred for the fits to the mid-IR. This suggests that CW2106 has a cloudy near-IR photosphere (primarily from deeper pressures of the atmosphere, $\sim 1-20$ bar), but a cloudless mid-IR photosphere (pressures $<1$), fitting the picture of an L/T transition object who's clouds have sedimented to the deeper layers of the atmosphere, but have not yet completely sunk below the observable photosphere. 

\subsection{Thermochemical Predictions}

In Section \ref{sec:2106_is_cloudy}, we discussed how the near-IR photosphere of CW2106 likely hosts silicate clouds. There are several different species of silicate clouds which can form in a brown dwarf atmosphere, each removing different amounts of material from the observable photosphere, most importantly oxygen. Atmospheric retrievals on T-dwarfs have found an excess of super-solar C/O ratios \citep[e.g.][]{Calamari_etal2022}, explained as the result of oxygen being sequestered into various condensate species \citep{Burrows1999,Visscher2010}. Being able to characterize which cloud species have formed in an atmosphere, and therefore how much oxygen has been sequestered by silicate condensates, will allow us to better constrain how much oxygen sequestration is taking place into other condensate species (for example into condensates of Ca, Ti, Al, and V).

\cite{Calamari_etal2024} outline a thermochemical framework which predicts the species of silicate clouds that can form given a bulk composition, finding the strongest predictor to be the ratio between Mg and Si (Mg/Si). For systems with a bulk Mg/Si $\gtrsim$0.9, MgSiO$_3$ is the dominant silicate species, with increasing contributions of Mg$_2$SiO$_4$ as Mg/Si increases, until Mg/Si$\gtrsim$1.6 when Mg$_2$SiO$_4$ becomes the dominant condensate. In atmospheres with Mg/Si$\sim$0.9, only MgSiO$_3$ is predicted to form, while atmospheres with Mg/Si$\lesssim$0.9 also have an increasing contribution from SiO$_2$ as Mg/Si decreases further.

In Section \ref{sec:CW2106_abunds} we derived an Mg/Si value for BD+24 of $\sim$0.87, which, as both objects form a dynamical pair, also applies to CW2106. This value sits just at the 0.9 threshold identified by \cite{Calamari_etal2024}, suggesting that the most likely cloud species to form in the atmosphere of CW2106 is MgSiO$_3$. Using this prediction, \cite{Calamari_etal2024} provide a way of estimating the amount of oxygen being sequestered, denoted as ``O$_{sink}$''. For CW2106, this value is O$_{sink}\approx23\%$, meaning that around 23\% of the bulk oxygen in CW2106 is now locked into condensate clouds. The bulk C/O of BD+24 (and thus CW2106) is $\approx0.63$, however, as 23\% of the oxygen is removed from the gas phase in CW2106, the atmospheric (i.e. observed) C/O ratio which is likely to be retrieved for CW2106 will be skewed to the higher value of $\sim0.82$ due to the clouds sequestering the Oxygen. Future retrievals will be able to affirm/refute this prediction, as well as shed more light on the possible presence of MgSiO$_3$ within it's atmosphere.

\section{CW2106 as a Benchmark} \label{sec:benchmarkTime!}

When attempting to estimate the parameters of isolated field dwarfs, it can be difficult to verify the trustworthiness of the results. To determine the validity of model parameters, it is necessary to compare the models with objects whose properties are already known. However, isolated brown dwarfs cannot be used to help improve model accuracy without ``ground truth'' values with which to correct to. CW2106, as part of a benchmark system, is one of the few brown dwarfs whose composition and age are known via its well-characterized host star. This information, combined with high SNR broadband spectroscopic data from JWST, means that CW2106 is well-poised to perform a direct test of the atmospheric models commonly used to estimate parameters of field objects. In Section \ref{sec:modeling}, we fit three different forward model grids to CW2106. Below we discuss some of the results of these model fits in context with the known parameters of CW2106 from its host star.

 \subsection{$T_{\rm eff}$}

The best fit overall for the full SED of CW2106 was the \texttt{Sonora Diamondback} model with $T_{\rm eff}$=1300 K. Both the \texttt{Elf-Owl} and \texttt{ATMO2020} model best fits also had $T_{\rm eff}$=1300 K. In Section \ref{sec:fund-params} we derived a $T_{\rm eff}$ for CW2106 of 1213$\pm21$ K using its semi-empirically measured $L_{\rm bol}$ along with an estimated radius using evolutionary models combined with the age of its host star. While similar, the $T_{\rm eff}$ of the models are $\sim$100 K hotter than that of CW2106, although the step size of the model grids is also 100 K.  

\subsection{$\log~g$}

The $\log~g$ values for the best fit models are all very low, with values of 3.5 for the \texttt{Diamondback}, \texttt{Elf Owl}, and \texttt{ATMO2020} models. These $\log~g$ values taken at face value would indicate that CW2106 is a very young brown dwarf. For example, using the evolutionary models of \cite{2008ApJ...689.1327S}, an object with a $T_{\rm eff}$ of $\sim$1200 K and $\log~g$=3.5 dex  would be $<10$ Myr old. Visually, the spectrum of CW2106 shows no obvious signatures of low gravity that would lead us to believe it is actually young, such as the triangular shaped H-band continuum shape \citep[e.g.,][]{Allers_etal2013}. Likewise, we know the age of the system to be 3.3$^{+0.7}_{-0.5}$ Gyr (Section \ref{sec:age}), much older than the model $\log~g$ values indicate. Using this age, we estimated the $\log~g$ in Section \ref{sec:fund-params} to be closer to $\sim$5.28 dex. Such a large mismatch in estimated surface gravity between this surface gravity and that from the best fitting model (\texttt{Sonora Diamondback}; 3.5 dex) is likely due to incomplete atmospheric chemistry and cloud physics in the model grids. 

\subsection{Metallicity}

BD+24, the host star of CW2106, has an overall metallicity of [M/H]=$0.00\pm0.06$ dex (Section \ref{sec:CW2106_abunds}). As its companion and substellar sibling, CW2106 should also have the same bulk metallicity. The best fitting model, \texttt{Sonora Diamondback}, has a metallicity of [M/H]=0 dex, in good agreement with the metallicity of BD+24. The best fitting \texttt{Sonora Elf Owl} model, however, has a supersolar metallicity of 0.5 dex. \texttt{Elf-Owl} does not include any contribution of clouds within its model grid, so the supersolar metallicity of the best fit model may be due to the higher metallicity more closely replicating the reddening effects caused by clouds.  

\subsection{CW2106 as a Synthetic Spectrum}

\begin{figure*}[ht]
    \centering
    \includegraphics[width=1.0\textwidth]{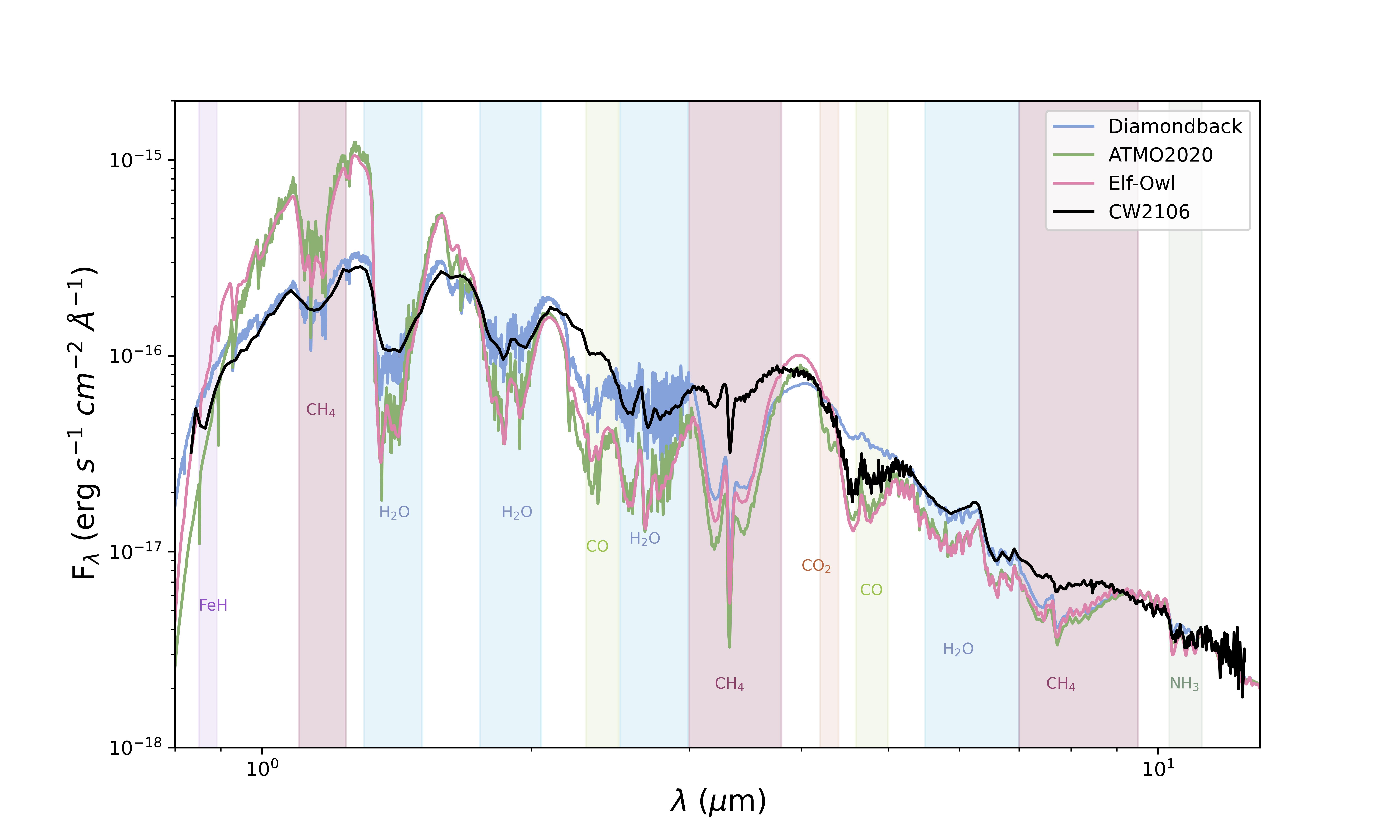}
    \caption{The SED of CW2106 (in black) along with synthetic SEDs from \texttt{Sonora Diamonback} (purple), \texttt{Sonora Elf-Owl} (pink), and \texttt{ATMO2020} (green). Regions of broad molecular absorptions are highlighted and labeled. All synthetic SEDs were generated for $T_{\rm eff}$=1213 K, $\log~g$=5.28 dex, and [M/H]=0.00 dex. The \texttt{Diamondback} SED was chosen to have $f_{sed}$=6 for the best visual fit, and both \texttt{Elf-Owl} \& \texttt{ATMO2020} SEDs have log(K$_{zz}$)=3 dex}.
    \label{fig:Synth_models}
\end{figure*}

Typically, fitting forward model grids to observed spectra is done as a way of inferring the parameters of your object. For isolated field brown dwarfs, there aren't many ways in which one can determine the plausibility of these parameters, leading one to rely on the ability of the models to accurately produce the spectra for a variety of different objects. In the case of CW2106, as discussed above, the parameters of the best-fit models in Section \ref{sec:fullm fit} do not match what we would expect from the analysis of its host star, which, if it were isolated, could possibly lead to an incorrect conclusion as to the nature of CW2106. 

While the model parameters provide a poor fit to those of CW2106, we can also ask another important question: how well do the models reproduce the SED of CW2106 using its \textit{actual} fundamental parameters? To answer this, we used \texttt{SEDA} to generate synthetic SEDs for each of the three models used in Section \ref{sec:modeling}, interpolating between model grid points to obtain models with the exact $T_{\rm eff}$, $\log~g$, and [M/H] of CW2106 (Section \ref{sec:fund-params}) and solar C/O (as all but the \texttt{Elf-Owl} models only consider solar C/O). For \texttt{Sonora Diamondback}, we did this for $f_{sed}$ values of 2, 4, 6, and 8, settling with a final value of $f_{sed}$=6 due to its best visual fit. Likewise, the $\log K_{zz}$ for both the \texttt{Elf-Owl} and \texttt{ATMO2020} were chosen via best visual fit as $\log K_{zz}$=3 dex.

The final synthetic SEDs are shown along with the observed SED of CW2106 in Figure \ref{fig:Synth_models}. Visually, \texttt{Diamondback} provides the best fit to the SED of CW2106 of all three synthetic SEDs used, although there are many regions where the models all fail to replicate the spectral features of CW2106. With the best fitting SED, from \texttt{Sonora Diamondback}, the model overpredicts the flux in the Y- and J-band. Likewise, \texttt{Diamondback} also over-predicts the amount of flux around $\sim$1.6 $\mu$m at the start of the H-band. Overall, \texttt{Diamondback} is able to replicate the depth of H$_2$O bands throughout the near-IR. However, at 2.2 $\mu$m, \texttt{Diamondback} has much stronger CH$_4$ absorption than is seen in CW2106. This is also true of the 3.3 $\mu$m CH$_4$ feature, which in the synthetic \texttt{Diamondback} SED is significantly deeper than that of CW2106. The 4 $\mu$m peak of the \texttt{Diamondback} SED is much sharper than that of CW2106, lacks the 4.2 $\mu$m CO$_2$ absorption feature, and is underluminous. The 4.6 $\mu$m CO feature, while present in the \texttt{Diamondback} SED, is considerably shallower than is actually observed, leading to an excess of flux in that region. In the mid-IR, while the depth of the H$_2$O in \texttt{Diamondback} is comparable to CW2106, the CH$_4$ is once again over-predicted, with a much stronger 7--9 $\mu$m feature in the \texttt{Diamondback} SED. The 10.5 $\mu$m NH$_3$ feature observed in CW2106 is also reproduced by the \texttt{Diamondback} model. 

The \texttt{Elf-Owl} and \texttt{ATMO2020} synthetic SEDs perform similarly to the \texttt{Diamondback} model, with stronger CH$_4$ features and overluminous J- and H-bands. However, these SEDs also have much deeper H$_2$O across the full wavelength range, likely because both the \texttt{Elf-Owl} and \texttt{ATMO2020} lack clouds, which have been shown to weaken these features \citep{Morley_etal2024}. Unlike the SED generated with the \texttt{Diamondback} models, the \texttt{Elf-Owl} and \texttt{ATMO2020} SEDs are able to accurately reproduced the observed 4.6 $\mu$m CO band depth, possibly due to the addition of disequilibrium chemistry within these models, unlike with the \texttt{Diamondback} models. The 4.2 $\mu$m CO$_2$ absorption feature in the \texttt{Elf-Owl} is weaker than observed in CW2106, while the CO$_2$ in \texttt{ATMO2020} appears to be stronger than is observed. Previous studies have also found an under-prediction of CO$_2$ using the \texttt{Elf-Owl} models.

\section{Conclusions} \label{sec:conclusions}
In this paper, we present the 0.8--12.5 $\mu$m SED, obtained with JWST and Magellan/FIRE, for the T0$\pm$1 dwarf CWISE J210640.16+250729.0, a companion to the K6V dwarf BD+24 4329. Using high resolution (R$\sim$60,000) spectroscopy of the host star, we derive elemental abundances for the system, finding it to be solar ([M/H]=0.00 dex) with an average Mg/Si ratio (Mg/Si=0.87). We derive a system age of 3.3$^{+0.7}_{-0.5}$ Gyr using model isochrones and gyrochronolgy, and fit its galactic orbit using Gaia DR3 kinematics, placing it within the thin disc population.

We semi-empirically derived the mass (M), radius (R), surface gravity ($\log~g$), effective temperature ($T_{\rm eff}$), and bolometric luminosity ($L_{\rm bol}$) for CW2106 using its full SED along with the age and distance adopted from its host star, finding values of 50--62 M$_{Jup}$, 0.83--0.87 R$_{Jup}$, 5.21--5.35 dex, 1213$\pm21$ K, and -4.825$\pm$0.005 \(\textup{L}_\odot\) respectively. Comparing these with literature values for other brown dwarfs, we find CW2106 is normal for its spectral type in $T_{\rm eff}$, but under-luminous compared with other objects of a similar temperature. We then presented an analysis of the spectral features observed in the JWST spectra, visually inspecting and calculating spectral indices of H$_2$O, CH$_4$, NH$_3$, and silicate condensates, placing these measurements into context with the broader field population, highlighting the ability of JWST to drastically reduce the errors associated with these calculations. 

We fit a series of different atmospheric model grids to the SED of CW2106, and find that the best fitting models require the inclusion of clouds, despite the lack of a silicate feature in the MIRI LRS spectrum. Fitting only portions of the SED provides inconsistent best fit model parameters depending on the region of choice. Thanks to the benchmark nature of CW2106, we are able to test the ability of current atmospheric models to accurately reproduce observed data. We find that all models prefer hotter $T_{\rm eff}$ values and lower $\log~g$ than was independently derived for CW2106, while models generated using the actual parameters of CW2106 fail to match its spectral morphology. 

While the atmospheric models are poor fits to the data, the preference for cloudy models, particularly in the near-IR (which probes the deepest atmospheric pressures), and the red color of CW2106, leads us to conclude that deep clouds in its atmosphere are likely helping to shape the SED of CW2106. This modeling suggests a cloudy near-IR photosphere, but a cloudless mid-IR photosphere. Using updated thermochemical models, as well as the Mg/Si ratio of the host star, we find that these clouds are expected to be comprised of MgSiO$_3$. Future atmospheric retrievals may be able to identify and confirm the species present \citep[e.g.,][]{Burningham_etal2021, Vos_etal2023}. 

This study highlights the immense power of using JWST's high signal, broad wavelength coverage observations with benchmark brown dwarfs. These JWST benchmarks will serve as a "Rosetta Stone", helping to decode and understand the roles that age, composition, and gravity play in sculpting sub-stellar spectra.

\begin{acknowledgments}
We thank Johanna Teske for helpful conversations which improved the paper. This paper includes data taken at The McDonald Observatory of The University of Texas at Austin. This paper includes data gathered with the 6.5 meter Magellan Telescopes located at Las Campanas Observatory, Chile. 
BB acknowledges support from UK Research and Innovation Science and Technology Facilities Council [ST/X001091/1].
VK acknowledges support from a UK Science and Technology Facilities Council studentship. JF and AR acknowledge support from NASA award 80NSSC22K0491, STSCI award JWST-GO-03670.002-A, and NSF award 2238468. J. M. V acknowledges funding from European Research Council Starting Grant Exo-PEA (Grant agreement No. [101164652]). This work presents results from the European Space Agency (ESA) space mission Gaia. Gaia data are being processed by the Gaia Data Processing and Analysis Consortium (DPAC). Funding for the DPAC is provided by national institutions, in particular the institutions participating in the Gaia MultiLateral Agreement (MLA). The Gaia mission website is https://www.cosmos.esa.int/gaia. The Gaia archive website is https://archives.esac.esa.int/gaia. This publication makes use of data products from the Wide-field Infrared Survey Explorer, which is a joint project of the University of California, Los Angeles, and the Jet Propulsion Laboratory/California Institute of Technology, funded by the National Aeronautics and Space Administration.
\end{acknowledgments}

% \begin{contribution}
% Author Rothermich obtained the data and processed it. X also wrote and edited the manuscript.
% Author Y provided the theoretical models and edited the manuscript.
% Author 1,2,3  contributed to the the manuscript.
% \end{contribution}

\facilities{JWST, Harlan J. Smith Telescope, Magellan Baade}

\software{Sedkit \citep{2020ascl.soft11014F}, JWST Science Calibration Pipeline \citep{bushouse_2022_7229890}, \texttt{BACCHUS} \citep{2016ascl.soft05004M}, \texttt{galpy} \citep{2015ApJS..216...29B}, \texttt{stardate} \citep{2019AJ....158..173A}, \texttt{SEDA} (\citealt{Suarez_etal2021}, Suarez et al. in prep), astropy, matplotlib}

\appendix

\section{Differential Brewer Star Comparison} \label{sec:brewer_results}

HD 131582 and HD 144872 were observed as part of the \cite{Brewer_cat} planet-search stars spectral catalog (hereinafter the Brewer catalog). All stars were observed with the HIRES spectrograph (R$\sim$70,000) located on the Keck I telescope over a variety of different radial velocity based planet searches. The spectra were uniformly analyzed using a procedure based on the Spectroscopy Made Easy (SME; \citealt{Valenti1996}) package, obtaining precise fundamental parameters (e.g., $T_{\rm eff}$, $\log~g$, vsin(i), etc.) and abundances for 15 elements using the Vienna Atomic Line Database (VALD-3).

HD 131582 and HD 144872 were used as standards in our differential abundance analysis of BD+24 (Section \ref{sec:bacchus}), using their parameters from \cite{Brewer_cat} along with their observed spectra (\ref{sec:tull_spec}) to calculate BD+24's offsets in each parameter from both standards. However, we also used this methodology to get the differential abundances of HD 131582 with respect to HD 144872, and vise versa. Comparing the results we get from this analysis with the ground truth values from \cite{Brewer_cat} provides a way of testing the performance of this analysis.      

The parameters of HD 131582 and HD 144872 from both \texttt{BACCHUS} and \cite{Brewer_cat} are listed in Table \ref{tab:Brewer_abunds}. Overall, our results using \texttt{BACCHUS} are in good agreement with those of the Brewer catalog. The \texttt{BACCHUS} $T_{\rm eff}$'s for HD 131582 and HD 144872 are consistent within $\lesssim1.7\sigma$ of those derived in \cite{Brewer_cat}. Likewise, the \texttt{BACCHUS} values for $\log~g$ and [M/H] are consistent within $\lesssim1\sigma$ and $\lesssim1.2\sigma$ respectively. 

\begin{figure*}[ht!]
    \centering
    \includegraphics[width=1.0\textwidth]{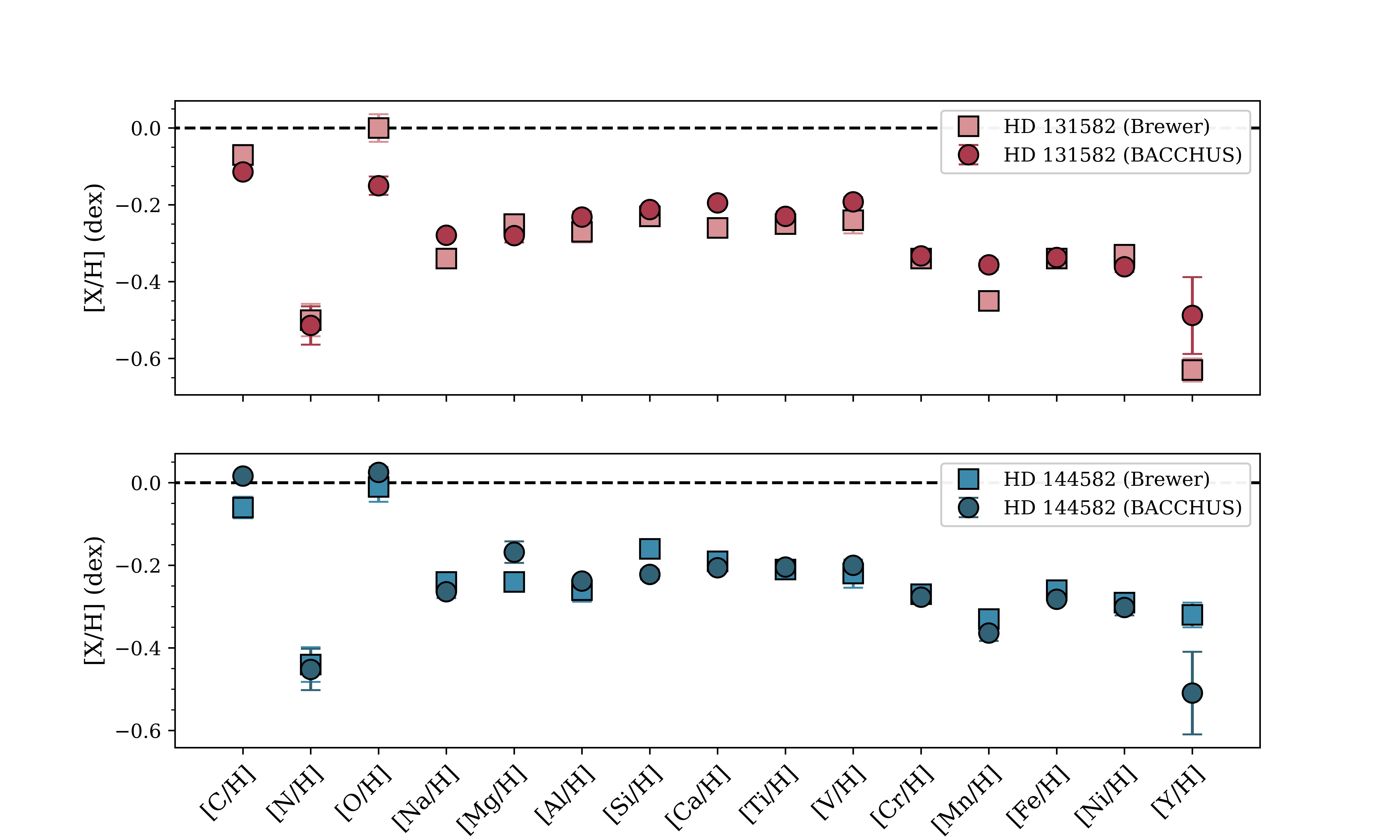}
    \caption{\textbf{Top:} Elemental abundances for HD 131582 from \texttt{BACCHUS} (circles) and from \cite{Brewer_cat} (squares) and their associated errors. The dashed line represents the reference Solar value of 0. \textbf{Bottom:} Same as in the top panel, but for HD 144872.}
    \label{fig:brewer_abund_comp}
\end{figure*}

Comparing the individual elemental abundances (Figure \ref{fig:brewer_abund_comp}) between our work and that of \cite{Brewer_cat} shows that the majority of elements for both HD 131582 and HD 144872 are consistent within $\lesssim2\sigma$. There are some noticeable departures, however. With HD 131582, for example, the [O/H] values derived by \cite{Brewer_cat} and those derived in this study differ by $\sim3.4\sigma$, with [O/H]$_{Brewer}=0.00\pm0.04$ and [O/H]$_{\texttt{BACCHUS}}=-0.15\pm0.02$. [Mn/H] is also slightly different, The results for [Ca/H] and [Na/H] in HD 133582 are each discrepant by $\sim6\sigma$. This difference, however, seems to be driven by the very small uncertainties in our \texttt{BACCHUS} values ($<0.01$ dex), rather than substantial differences between the two results. Due to the slight differences in individual elemental abundances, the C/O and Mg/Si ratios of HD 131582 are a bit discrepant, with differences of $2.9\sigma$ and $4.9\sigma$ respectively. 

The \texttt{BACCHUS} individual abundances of HD 144872 have an overall stronger agreement with \cite{Brewer_cat} than was seen in HD 131582. However, one major difference is between the measured [Si/H] which differ by $\sim4.2\sigma$. While both \cite{Brewer_cat} and our study find subsolar values of [Si/H] for HD 144872, our results show a slightly more subsolar value. As a result, the Mg/Si values for HD 144872 differ substantially ($\sim9.5\sigma$). 
 
One important caveat is that these differential abundances for HD 131582 and HD 144872 were derived using only one reference star from which their difference were calculated. Ideally, these differences would be calculated with multiple reference stars as was done in Section \ref{sec:bacchus} for BD+24. Future works will expand on these differential abundances using more sources from \cite{Brewer_cat}, allowing for a proper benchmarking of the technique.

 \begin{deluxetable*}{lcccc}
    \tabletypesize{\scriptsize}
    \tablewidth{0pt}
    \tablenum{3}
    \tablecolumns{5}
    \tablecaption{Differential Abundance Comparison.}\label{tab:Brewer_abunds}
    \tablehead{ 
    \multicolumn{1}{c}{} & \multicolumn{2}{c}{HD 131582} & \multicolumn{2}{c}{HD 144872} \\
    \cline{2-3}
    \cline{4-5}
    \colhead{Property} & 
    \colhead{Lit.$^a$}&
    \colhead{\texttt{BACCHUS$^b$}}&
    \colhead{Lit.$^a$} & 
    \colhead{\texttt{BACCHUS$^c$}}}
    \startdata
        $T_{\rm eff}$ (K) & 4728$\pm$25 & 4772$\pm$8 & 4780$\pm$25 & 4814$\pm$32 \\\relax
        $\log~g$ (dex)  & 4.52$\pm$0.03 & 4.58$\pm$0.54 & 4.52$\pm$0.03 & 4.58$\pm$0.40 \\\relax
        [M/H] (dex)  & -0.29$\pm$0.01 & -0.31$\pm$0.08 & -0.26$\pm$0.01 & -0.20$\pm$0.05 \\\relax
        [Fe/H] (dex)  & -0.34$\pm$0.01 & -0.34$\pm$0.01 & -0.26$\pm$0.01 & -0.28$\pm$0.01 \\\relax
        [C/H] (dex)  & -0.07$\pm$0.03 & -0.11$\pm$0.01 & -0.06$\pm$0.03 & 0.02$\pm$0.01 \\\relax
        [N/H] (dex)  & -0.50$\pm$0.03 & -0.51$\pm$0.05 & -0.44$\pm$0.03 & -0.45$\pm$0.05 \\\relax
        [O/H] (dex)  & 0.00$\pm$0.04 & -0.15$\pm$0.02 & -0.01$\pm$0.04 & 0.03$\pm$0.01 \\\relax
        [Mg/H] (dex)  & -0.25$\pm$0.01 & -0.28$\pm$0.02 & -0.24$\pm$0.01 & -0.17$\pm$0.03 \\\relax
        [Si/H] (dex)  & -0.23$\pm$0.01 & -0.21$\pm$0.01 & -0.16$\pm$0.01 & -0.22$\pm$0.01 \\\relax
        [Ca/H] (dex)  & -0.26$\pm$0.01 & -0.20$\pm$0.01 & -0.19$\pm$0.01 & -0.21$\pm$0.01 \\\relax
        [Na/H] (dex)  & -0.34$\pm$0.01 & -0.28$\pm$0.01 & -0.24$\pm$0.01 & -0.26$\pm$0.02 \\\relax
        [Al/H] (dex)  & -0.27$\pm$0.03 & -0.23$\pm$0.02 & -0.26$\pm$0.03 & -0.24$\pm$0.01 \\\relax
        [Ti/H] (dex)  & -0.25$\pm$0.01 & -0.23$\pm$0.01 & -0.21$\pm$0.01 & -0.20$\pm$0.01 \\\relax
        [V/H] (dex)  & -0.24$\pm$0.03 & -0.19$\pm$0.01 & -0.22$\pm$0.03 & -0.20$\pm$0.01 \\\relax
        [Cr/H] (dex)  & -0.34$\pm$0.01 & -0.33$\pm$0.01 & -0.27$\pm$0.01 & -0.28$\pm$0.01 \\\relax
        [Mn/H] (dex)  & -0.45$\pm$0.02 & -0.36$\pm$0.01 & -0.33$\pm$0.02 & -0.36$\pm$0.02 \\\relax
        [Ni/H] (dex)  & -0.33$\pm$0.01 & -0.36$\pm$0.01 & -0.29$\pm$0.01 & -0.30$\pm$0.02 \\\relax
        [Y/H] (dex)  & -0.63$\pm$0.03 & -0.49$\pm$0.10 & -0.32$\pm$0.03 & -0.51$\pm$0.10 \\\relax
        C/O  & 0.47$\pm$0.04$^*$ & 0.60$\pm$0.02 & 0.49$\pm$0.04$^*$ & 0.54$\pm$0.01 \\\relax
        Mg/Si  & 0.98$\pm$0.01$^*$ & 0.87$\pm$0.02 & 0.85$\pm$0.01$^*$ & 1.15$\pm$0.03 \\
    \enddata
    \tablecomments{a - Literature values from \cite{Brewer_cat}.\newline
    b - \texttt{BACCHUS} derived differential abundances with respect to HD 144872. \newline
    c - \texttt{BACCHUS} derived differential abundances with respect to HD 131582. \newline
    * - This value was derived in this study from the abundances presented in \cite{Brewer_cat}.}
\end{deluxetable*}

\section{Raw Differential Abundances for BD+24} \label{sec:fullDifferences}

As discussed in Section \ref{sec:bacchus}, the abundances for BD+24 presented in Section \ref{sec:brewer_results} are derived by comparing the spectra, line by line, between BD+24 and the reference stars HD 131582 and HD 144872, determining the difference in absolute abundance between each star. These differences are then applied to the final abundance results presented in \cite{Brewer_cat} to get the corresponding ``absolute'' abundances for BD+24. However, this method limits the analysis to the 15 elements investigated by \cite{Brewer_cat}, as these are the only values to which we can apply BD+24's offsets. Using \texttt{BACCHUS} with the broadband wavelength coverage spectra obtained with the Tull Coude Spectrograph (Section \ref{sec:tull_spec}), our analysis is sensitive to 26 individual elements. Table \ref{tab:raw_differences} presents the raw differential abundances (i.e. $\Delta$[X/H]) for BD+24 with respect to both HD 131582 and HD 144872 for all elements we are able to constrain using \texttt{BACCHUS}.  

 \begin{deluxetable*}{lcccc}
    \tabletypesize{\scriptsize}
    \tablewidth{0pt}
    \tablenum{4}
    \tablecolumns{3}
    \tablecaption{BD+24 Raw Differential Abundances.}\label{tab:raw_differences}
    \tablehead{ 
    \colhead{Property} & 
    \colhead{$\Delta_{131}^a$}&
    \colhead{$\Delta_{144}^b$} }
    \startdata
        $\Delta T_{\rm eff}$ (K)	&	-228.00	$\pm$	79.00	&	-268.00	$\pm$	21.00	\\\relax
        $\Delta \log~g$ (dex)	&	-0.13	$\pm$	0.05	&	-0.02	$\pm$	0.10	\\\relax
        $\Delta$[M/H] (dex)	&	0.29	$\pm$	0.26	&	0.26	$\pm$	0.06	\\\relax
        $\Delta$v$_{mic}$ (Km/s$^2$)	&	0.00	$\pm$	0.13	&	0.00	$\pm$	0.03	\\\relax
        $\Delta$[C/H] (dex)	&	0.24	$\pm$	0.02	&	0.22	$\pm$	0.02	\\\relax
        $\Delta$[N/H] (dex)	&	0.25	$\pm$	0.05	&	0.25	$\pm$	0.05	\\\relax
        $\Delta$[O/H] (dex)	&	0.11	$\pm$	0.08	&	0.10	$\pm$	0.09	\\\relax
        $\Delta$[Na/H] (dex)	&	0.20	$\pm$	0.10	&	---			\\\relax
        $\Delta$[Mg/H] (dex)	&	0.36	$\pm$	0.03	&	0.32	$\pm$	0.01	\\\relax
        $\Delta$[Al/H] (dex)	&	0.24	$\pm$	0.01	&	0.17	$\pm$	0.01	\\\relax
        $\Delta$[Si/H] (dex)	&	0.35	$\pm$	0.07	&	0.35	$\pm$	0.06	\\\relax
        $\Delta$[Ca/H] (dex)	&	0.29	$\pm$	0.03	&	0.19	$\pm$	0.03	\\\relax
        $\Delta$[Sc/H] (dex)	&	0.34	$\pm$	0.04	&	0.22	$\pm$	0.04	\\\relax
        $\Delta$[Cr/H] (dex)	&	0.26	$\pm$	0.01	&	0.18	$\pm$	0.01	\\\relax
        $\Delta$[Ti/H] (dex)	&	0.18	$\pm$	0.04	&	0.16	$\pm$	0.02	\\\relax
        $\Delta$[V/H] (dex)	&	0.22	$\pm$	0.10	&	0.15	$\pm$	0.01	\\\relax
        $\Delta$[Mn/H] (dex)	&	0.36	$\pm$	0.10	&	0.21	$\pm$	0.10	\\\relax
        $\Delta$[Co/H] (dex)	&	0.23	$\pm$	0.05	&	0.32	$\pm$	0.04	\\\relax
        $\Delta$[Fe/H] (dex)	&	0.30	$\pm$	0.01	&	0.25	$\pm$	0.01	\\\relax
        $\Delta$[Ni/H] (dex)	&	0.31	$\pm$	0.02	&	0.27	$\pm$	0.02	\\\relax
        $\Delta$[Cu/H] (dex)	&	0.26	$\pm$	0.10	&	0.26	$\pm$	0.10	\\\relax
        $\Delta$[Zn/H] (dex)	&	0.17	$\pm$	0.02	&	0.16	$\pm$	0.02	\\\relax
        $\Delta$[Y/H] (dex)	&	---			&	0.29	$\pm$	0.10	\\\relax
        $\Delta$[Zr/H] (dex)	&	0.40	$\pm$	0.02	&	0.19	$\pm$	0.03	\\\relax
        $\Delta$[Mo/H] (dex)	&	0.31	$\pm$	0.01	&	0.23	$\pm$	0.10	\\\relax
        $\Delta$[Ba/H] (dex)	&	0.30	$\pm$	0.10	&	0.02	$\pm$	0.10	\\\relax
        $\Delta$[La/H] (dex)	&	0.40	$\pm$	0.05	&	---			\\\relax
        $\Delta$[Nd/H] (dex)	&	0.26	$\pm$	0.10	&	---			\\\relax       
    \enddata
    \tablecomments{a - Differential values with respect to HD 131582. \newline
    b - Differential values with respect to HD 144872.}
\end{deluxetable*}

\FloatBarrier
\bibliography{references_2025}{}
\bibliographystyle{aasjournal}

\end{document}